\documentclass[twocolumn,aps,showpacs,prb,epsf,graphics,psfig]{revtex4-2}
\usepackage{graphicx}
\usepackage{graphicx}
\usepackage{bbold}
\usepackage{color}
\usepackage[normalem]{ulem}
\usepackage{amssymb}
\usepackage{tikz}
\usepackage[version=3]{mhchem} 
\usepackage{amsmath}
\usepackage[export]{adjustbox}

\usepackage{lineno}

\begin{document}
\title{A Monte Carlo simulation framework for investigating the effect of inter-track coupling on H$_2$O$_2$ productions at ultra-high dose rates}
\author{Ramin Abolfath$^{1,2,3\dagger}$, Sedigheh Fardirad$^{4}$,
Houda Kacem$^5$, Marie-Catherine Vozenin$^{5,6,7}$, and Abbas Ghasemizad$^{4}$}
\affiliation{
$^1$Department of Physics and Astronomy, Howard University, Washington, DC 20059, USA \\
$^2$Department of Radiation Physics and Oncology, University of Texas MD Anderson Cancer Center, Houston, TX, 75031, USA \\
$^3$Physics Department, Sharif University of Technology, P.O.Box 11365-9161, Azadi Avenue, Tehran, Iran \\
$^4$Department of Physics, Faculty of Science, University of Guilan, P.O. Box 41335-1914, Rasht, Iran \\
$^5$ The sector of radiobiology applied to radiotherapy/Radiation Oncology Department/Geneva  
University Hospital, Geneva, Switzerland.  \\
$^6$ LiRR- laboratory of innovation in radiobiology applied to radiotherapy/Faculty of 
Medicine/University of Geneva, Geneva, Switzerland. \\
$^7$ Laboratory of Radiation Oncology/Radiation Oncology Service/Department of Oncology 
/CHUV; Lausanne University Hospital and University of Lausanne, Lausanne, Switzerland. 
}
\date{\today}

\begin{abstract}
{\bf Background}:
Lower production of H$_2$O$_2$ in water is a hallmark of ultra-high dose rate (UHDR) compared to the conventional dose rate (CDR).
However, the current computational models based on the predicted yield of H$_2$O$_2$ are the opposite of the experimental data.

{\bf Purpose}:
To present a multi-scale formalism to reconcile the theoretical modeling and the experimental observations of H$_2$O$_2$ production and provide a mechanism for the suppression of H$_2$O$_2$ at FLASH-UHDR.

{\bf Methods}:
We construct an analytical model for the rate equation in the production of H$_2$O$_2$ from \ce{^{.}OH}-radicals and use it as a guide to propose a hypothetical geometrical inhomogeneity in the configuration of particles in the UHDR beams. We perform a series of Monte Carlo (MC) simulations of the track structures for a system of charged particles impinging the medium in the form of clusters and/or bunches.

{\bf Results}:
We demonstrate the interplay of diffusion, reaction rates, and overlaps in track-spacing attribute to a lower yield of H$_2$O$_2$ at UHDR vs. CDR. This trend is reversed if spacing among the tracks becomes larger than a critical value, with a length scale that is proportional to the diffusion length of \ce{^{.}OH}-radicals modulated by a rate of decay due to recombination with other species, available within a track, and the space among the tracks. The latter is substantial on the suppressing of the H$_2$O$_2$ population at UHDR relative to CDR. Under these conditions in our MC setup, the reduction of H$_2$O$_2$ dose rate effect within 1ms time-scale is attributed mainly to several orders of magnitude earlier times, e.g., 1 ns of the e$^-_{\rm aq}$ reaction with \ce{^{.}OH}-radicals and negligible reaction with \ce{^{.}H}.

{\bf Conclusions}:
Based on our analysis of the present work, at UHDR, the lower yield in H$_2$O$_2$ can be interpreted as a signature of bunching the particles in beams of ionizing radiation, and temporal correlations and time-dependent chain of reactions. The beams enter the medium in closely packed clusters and form inhomogeneities in the track-structure distribution. Thus the MC simulations based on the assumption of uniformly distributed tracks are unable to explain the experimental data.

\end{abstract}
\maketitle

\maketitle

\section{Introduction}
\label{Sec_Intro}
Recently an intensive interest in understanding the radio-biological effects of the FLASH ultra-high dose rate (UHDR) radiotherapy has emerged, experimentally and theoretically, due to
the observed unique normal tissue sparing of 40 Gy/s and higher
[\onlinecite{Favaudon2014:STM,Montay-Gruel2018:RO,Vozenin2018:CCR,Montay-Gruel2019:PNAS,Buonanno2019:RO,Vozenin2019:RO,Arash2020:MP,Spitz2019:RO,Koch2019:RO,Houda:PhDThesis,Kacem2024:SA,Zhang2024:MP,Abolfath2020:MP,Seco2021:MP,Abolfath2023:FP,Baikalov2023:FP,Blain2022:RR,Kacem2022:IJRB,Montay-Gruel2019:PNAS}].

A series of systematic experiments in water have been conducted to compare the yields in H$_2$O$_2$ production at two levels of low and high dose rates.
Accordingly, the experimental yield at FLASH-UHDR is less than that of conventional dose rates (CDR) [\onlinecite{Blain2022:RR,Kacem2022:IJRB,Montay-Gruel2019:PNAS,Houda:PhDThesis,Kacem2024:SA}].

Several research teams have examined the experimental data against the computation of the vastly used reaction rate models for chemical simulation of water radiolysis [\onlinecite{chemsimul}], in addition to the Monte Carlo (MC) computational models [\onlinecite{Lai2021:PMB,Thompson2023:PMB,Derksen2023:PMB}].
Accordingly, it has turned out that the result of their computational models has predicted contradictory outcomes, i.e., the model calculations have predicted the higher yield of H$_2$O$_2$ at FLASH dose rates, compared to CDR [\onlinecite{Houda:PhDThesis,Kacem2024:SA,Zhang2024:MP}].

The present study aims to suggest a theoretical model combined with a Monte Carlo track structure to reconcile the theoretical predictions with the experiments.
Following the recent models proposed by the first author [\onlinecite{Abolfath2020:MP}], in this work, we demonstrate analytically and validate it numerically how the inter-track coupling may be responsible for the lower yield of H$_2$O$_2$ at UHDR.

We parameterize the dose rate by temporal and spatial spacing among the tracks
[\onlinecite{Abolfath2020:MP,Abolfath2023:FP,Baikalov2023:FP}].
We show that the yield in the production of H$_2$O$_2$ by two \ce{^{.}OH}-radicals increases as track spacing decreases up to a characteristic length where H$_2$O$_2$ production saturates and falls to a lower value asymptotically.
The specific length is proportional to the effective diffusion length of \ce{^{.}OH}-radicals, that is the mean free path modulated by the scavenging rate.
Finally, a recent multi-tracks simulation in TRAX-CHEM [\onlinecite{Castelli2025:IJMS}] has shown agreement with our models and the results presented in an earlier version of the present work [\onlinecite{Abolfath2024:ArXiV}].


%
%
\section{Methods and model calculations}
\label{Sec_Method}
Our methods to investigate the effects of inter-track separation on the H$_2$O$_2$ yield consist of analytical derivations and testing the predictions by performing a series of experimental Geant4-DNA MC simulations of the track structure.
To this end, we used the chemistry and UHDR modules of Geant4-DNA [\onlinecite{Geant4DNA1,Geant4DNA2,Geant4DNA3,Geant4DNA4,Geant4DNA5}].

We calculate the yield in H$_2$O$_2$ production based on a non-Poissonian distribution of the absorbed dose to demonstrate the scavenging rate of \ce{^{.}OH}-radicals is an increasing function of the tracks compactness.
We isolate the production of H$_2$O$_2$ by \ce{^{.}OH}-radicals from all other possible chemical reactions, including the recombination of \ce{^{.}OH}-radicals with \ce{^{.}H} and e$^-_{\rm aq}$.
In the first part, we take advantage of this simplification and reduce the rate equations to a single non-linear differential equation. In MC, we, however, use the entire rate equations available in the chemistry and UHDR modules of Geant4-DNA.

The UHDR extended/medical/DNA example illustrates how to activate the chemistry mesoscopic model in combination with the step-by-step energy transfer/deposition model. It allows the simulation of chemical reactions beyond 1 $\mu$s post-irradiation. 
The UHDR example in the Geant4-DNA toolkit can utilize both the track structure (or event-by-event) and the mesoscopic simulation approaches.

In Geant4-DNA, the step-by-step track structure simulation (stochastic boundary simulation) involves (a) the explicit simulation of the transport and interactions of individual particles (e.g., electrons, protons, $\cdots$) on an event-by-event basis, and (b) a detailed registration of the physical and chemical processes at the nanometer scale, capturing the stochastic nature of radiation interactions.
This approach is particularly useful for studying the initial physical stage of radiation interactions and the production of reactive species within the cellular environment.

The mesoscopic simulation approach in Geant4-DNA aims to bridge the gap between the detailed track structure simulation and the macroscopic continuum models.
It uses a coarse-grained representation of the cellular and sub-cellular structures, allowing for the simulation of larger spatial and temporal scales compared to the track structure simulation.
The mesoscopic approach can be used to model the diffusion and chemical reactions of the radiation-induced reactive species, as well as their potential impact on cellular structures and functions.
For instance, the initial physical stage of the radiation interactions may be simulated using the track structure approach to accurately capture the production and spatial distribution of the reactive species. Then, the mesoscopic simulation can be used to model the subsequent diffusion and chemical reactions of these species within the cellular environment, and their potential impact on cellular structures and processes.
However, there have not been molecular dynamics in the underlying cellular structure of the medium. 

The UHDR example uses the Geant4-DNA chemistry extension, which provides a set of models and processes for simulating the chemical reactions that occur after the initial physical interactions of radiation with biological materials. This chemistry list includes a wide range of reaction rate coefficients and cross-sections that have been validated against experimental data (under normal conditions). This allows for a detailed simulation of the complex radiation chemistry processes that occur within the cellular environment. Therefore, with some limitations on the input parameters, this set of computational packages can be employed for modeling the spatial and temporal evolution of the radiation-induced reactive species and their potential impact on cellular structures and functions under ultra-high dose rate conditions.
It starts with the radiolysis of water including (a) ionization and excitation of water molecules, (b) production of reactive species such as hydroxyl radicals (\ce{^{.}OH}), hydrated electrons ($e^-_{\rm aq}$), and hydrogen atoms (\ce{^{.}H}). The diffusion and chemical reactions of the reactive species include (a) diffusion of the species based on their respective diffusion coefficients under ambient conditions (no additional boost due to the formation of thermal spikes) and (b) bimolecular and termolecular chemical reactions between the various species. 

The implementation of oxygen concentration that is available in the Geant4-DNA-UHDR example, is typically set as a parameter that can be adjusted by the user in the simulation setup. This allows for the investigation of the impact of different oxygen levels on the radiation-induced chemical processes and their consequences.
The general implementation of oxygen concentration in Geant4-DNA can be summarized as follows (by defining it in source and macro files):
(a) The initial oxygen concentration is defined as a parameter in the simulation setup, typically in units of molarity (M) and density; by defining water based on NIST material database.
(b) During the simulation, the transport and diffusion of oxygen molecules within the cellular environment are modeled using appropriate diffusion coefficients.
(c) The Geant4-DNA toolkit tracks the spatial and temporal changes in oxygen concentration, allowing for the investigation of hypoxic conditions in a homogeneous underlying medium and their impact on the radiation-induced chemical processes and biological effects.

Finally, to avoid any misunderstanding of the present study, we must point out that the correct interpretation of our study is the following: if the users of MC codes, e.g., Geant4-DNA or TOPAS change their initial conditions from a uniform distribution of particles to a specific initial condition and sample the charged particles initiated from a source in which selective clusters and bunches of particles can be selected, the MC codes should correctly give lower G-values for H$_2$O$_2$ at UHDR relative to CDR.
\subsection{Reaction rate equations of a system of many tracks - a Markov chain}
Let us consider at time $t$, after radiation,  a large number of reactive species (RS), e.g., \ce{^{.}OH}-radicals, \ce{^{.}H}-radicals, $e^-_{\rm aq}$, $\cdots$, have been created.
They move out of the center of creation randomly, i.e., by boosted Brownian motion and thermal diffusion, thermal spikes, shock waves, etc.
Once they meet each other within a certain distance, in spatial proximity to each other, they react chemically and form either metastable or short-lifetime transient complexes of
ranked-$r$ clusters
or stable non-reactive species (NRS or larger compounds), such as H$_2$O$_2$ with 
$r=2$.
We call these complexes, NRS, because of their larger mass and lower mobility compared to RS to reach out to bio-molecules and form any types of damage [\onlinecite{Abolfath2020:MP}].
The dimension of these clusters, $\ell_c$, are assumed to be comparable with the diffusive mean free path of \ce{^{.}OH}-radicals, $\ell = \sqrt{2D_f\tau}$, where $D_f$ and $\tau$ are the diffusion constant and the relaxation time, corrected for the scavenging rate of \ce{^{.}OH}-radicals.

For the simplicity in modeling, we focus on the dynamics of \ce{^{.}OH}-radicals.
The following rate equation, derived from a stochastic Markov chain Master equation, governs the reaction rate of $n$-\ce{^{.}OH}-radicals
\begin{eqnarray}
\frac{d\overline{n}(t)}{dt} &=& \mu \dot{z} - \gamma_1 \overline{n} - \frac{\gamma_2}{2!}\overline{n(n-1)}
- \frac{\gamma_3}{3!}\overline{n(n-1)(n-2)} \nonumber \\
&-& ... - \frac{\gamma_N}{N!}\overline{n(n-1)(n-2)...(n-N)},
\label{eq1}
\end{eqnarray}
where $N=r-1$.
The first term describes the rate of energy deposition by charged particles.
$\mu$ is a constant, describing the radiation yield of \ce{^{.}OH}-radicals per unit of dose.  
The other constants, $\gamma_1, \gamma_2, \cdots$ are reaction rate constants. 
They are the intra-track dependent parameters, sensitive to the spatial compactness of local ionization density profile and hence the single particle LET.
They are independent of the spatial distribution of the tracks. 
Thus they describe decay rates of \ce{^{.}OH}-radicals in a system of non-interacting tracks.
The dependencies of $n$ on the temporal and spatial distribution of the tracks, hence inter-track correlations, stems from the rest of the $n$-combinatorial factors (see below) and 
$\dot{z}$, the dose rate. Here $z=\varepsilon/m$ is the specific energy (energy divided by mass) or the absorbed dose.
$\varepsilon$ is the energy, deposited in mass $m$.
The second term, linear in $\overline{n}$, describes the scavenging or decay rate of \ce{^{.}OH}-radicals.
The over-line is the statistical averaging.
The third term describes the formation of a cluster of two out of $n$ \ce{^{.}OH}-radicals.
Similarly, the rest describe clustering of three, four, ..., and $N$ \ce{^{.}OH}-radicals.
The factorial numbers, $2!, 3!, \cdots, N!$ account for the combinatorial factors of clustering $N$ objects out of $n$-identical objects.
Out of these clusters, $\overline{n(n-1)}/2$ combinations, averaged over an appropriate distribution function (see below), gives the population of created H$_2$O$_2$ out of $n$ \ce{^{.}OH}-radicals.

The statistical averaging over different moments of the clusters,
$\overline{n(n-1)(n-2)...(n-N)}$, can be performed by integrating over a compound Poisson or Neyman's distribution of type A that governs the probability distribution of 
absorbed dose [\onlinecite{Virsik1981:RR},\onlinecite{Gudowska2000:APP},\onlinecite{Abolfath2019:EPJD}]
\begin{eqnarray}
Q_n(\theta, \Delta) = \sum_{\nu=0}^\infty P_\nu(\theta) P_n(\nu\Delta).
\label{eq_Qn}
\end{eqnarray}
Here $\nu$ is the number of tracks passing through the volume of interest.
$\theta = \overline{\nu}$ is proportional to the average number of tracks that give the mean absorbed dose.
For a given dose, the mean number of tracks, $\theta$, is assumed to be statistically a given number, independent of the dose rate (the time lag among tracks).
In the limit of large number of tracks, the same implication can be deduced for $\nu$, e.g., the same integer number of tracks lead to a given absorbed dose, independent of their relative time lags or rate of entrance of each single track.  
In a system of many tracks,
$n$ is the number of collisions (distributed over all tracks) that lead to the creation of \ce{^{.}OH}-radicals out of $M$ initial ionization and molecular excitations leading to the creation of all RS. 
$\Delta$ describes the spatial compactness of all (intra- and inter-track) ionizations and excitations and is a radiation characteristic parameter.  
In Eq. (\ref{eq_Qn}), we have factorized $\Delta$ from $\nu$, the number of tracks. 
Thus $\Delta$ times $\overline{\nu}$ is proportional to the average number of $n$ \ce{^{.}OH}-radicals generated by all tracks. 
This dependence should be visible from Poisson distribution, $P_n(\nu\Delta)$, in Eq. (\ref{eq_Qn}).
Thus $\Delta$ contains both intra- and inter-track effects of the radiation.  
Here a track and event are interchangeable.
Note that $M>n$. Thus $Q_n(\theta, \Delta)$ is the reduced density matrix, in a sense that all other RS degrees of freedom (\ce{^{.}H}-radicals, $e^-_{\rm aq}$, $\cdots$) have been traced/integrated out, $Q_n = {\rm Tr}_M Q_M$.
Performing averaging over the distribution given by Eq. (\ref{eq_Qn}) results in $\overline{n} = \sum_{n=0}^\infty n Q_n(\theta, \Delta) = \theta \Delta$.
From this dependence it must be clear that $\theta$ and $\Delta$ are the mean track and the spatial compactness of \ce{^{.}OH}-radicals. 
Similarly, $\overline{n(n-1)} = (\theta^2 + \theta) \Delta^2 = \overline{n}^2 + \overline{n}\Delta$,
$\overline{n(n-1)(n-2)} = (\theta^3 + 2\theta^2 + \theta) \Delta^3 = \overline{n}^3 + 2\overline{n}^2\Delta + \overline{n} \Delta^2$,
$\overline{n(n-1)(n-2)(n-3)} = (\theta^4 + 6 \theta^3 + 7\theta^2 + \theta) \Delta^4 = \overline{n}^4 + 6\overline{n}^3\Delta + 7 \overline{n}^2 \Delta^2 + \overline{n}\Delta^3$,
$\cdots$.

After a tedious but straightforward algebra, the rate equation for $\overline{n}$ can be derived
\begin{eqnarray}
\frac{d\overline{n}(t)}{dt} &=& \mu \dot{z}
- (\gamma_1+\frac{\gamma_2}{2!}\Delta+\frac{\gamma_3}{3!}\Delta^2+\frac{\gamma_4}{4!}\Delta^3 + \cdots) \overline{n}
\nonumber \\
&-& (\frac{\gamma_2}{2!}\Delta+2\frac{\gamma_3}{3!}\Delta^2+7\frac{\gamma_4}{4!}\Delta^3 + \cdots) \overline{n}^2
\nonumber \\
&-& (\frac{\gamma_3}{3!}\Delta^2 + 6\frac{\gamma_4}{4!}\Delta^3 + \cdots) \overline{n}^3 \nonumber \\
&-& (\frac{\gamma_4}{4!}\Delta^3 +\cdots) \overline{n}^4 - \dots
\label{eq4}
\end{eqnarray}
In this form of rate equation, $\overline{n}$ decays as a function of time as \ce{^{.}OH}-radicals evolve into other types of species, and their tracks fade away.
The higher $\Delta$, the faster decay in $\overline{n}$, i.e., the faster-emerging rate of individual tracks into a single cloud.  

Truncating the series in the rate equation beyond the quartic terms in $\overline{n}$ gives
\begin{eqnarray}
\frac{d\overline{n}(t)}{dt} &=& \mu \dot{z} - \lambda_{\rm eff}(\Delta) \overline{n} - \gamma_{\rm eff}(\Delta) \overline{n}^2 - \eta _{\rm eff}(\Delta) \overline{n}^3
- {\cal O}(\overline{n}^4), \nonumber \\
\label{Eq_rate_1}
\end{eqnarray}
where
\begin{eqnarray}
\lambda_{\rm eff}(\Delta) = \gamma_1+\frac{\gamma_2}{2!}\Delta+\frac{\gamma_3}{3!}\Delta^2+\frac{\gamma_4}{4!}\Delta^3 + \cdots
\label{Eq_rate_2}
\end{eqnarray}
and
\begin{eqnarray}
\gamma_{\rm eff}(\Delta) = \frac{\gamma_2}{2!}\Delta+2\frac{\gamma_3}{3!}\Delta^2+7\frac{\gamma_4}{4!}\Delta^3 + \cdots
\label{Eq_rate_3}
\end{eqnarray}
are the polynomials of $\Delta$. The same for $\eta _{\rm eff}(\Delta)$.

In these equations, the parameter $\Delta$ that has been introduced in the energy deposition distribution function corrects the scavenging rates of RS due to the intra- and inter-track recombination of RS. 
As $\Delta$ depends on the inter-track spacing and the spatial compactness of the ionization and excitation in space at a given moment, its value depends on the temporal distribution of the tracks, which is given at a specific dose rate.
Note that for a given dose rate, and given the inter-track separation, $\Delta$ is a constant of time.
It might also be useful to recall a close analogy for $\Delta$ with the change of charged particles LET as a function of depth in a medium in a limit of large inter-track spacing or CDR where a system of many tracks can be reduced to a single track (with no inter-track interaction). 
In a system of non-interacting tracks, the response of the system increases monotonically by increasing $\Delta$ (single-particle LET) until it drops at a saturation point where a molecular (over) crowding of chemical species takes place. Such saturation has been discussed in the context of DNA damage and radiobiological effectiveness of radiation, RBE, as a function of LET, where the turning point has referred to a so-called overkilling effect in RBE response of cells (a misleading terminology).
For the rest of this study, we hypothesize under a variation of the dose rate, e.g., switching the source of radiation from CDR to FLASH-UHDR, the local compactness of RS increases so the numerical values of $\Delta$ jump up. 

To investigate further the connection between $\Delta$ and track-spacing, we start from the relation $\Delta = \mu z_D$ where $z_D = (\overline{z^2}-
\overline{z}^2)/\overline{z}$ is the renormalized fluctuations of $z$ (for the derivation, see Ref. [\onlinecite{Abolfath2019:EPJD}]).
Let us consider two tracks with labels 1 and 2 with specific energies $z_1$ and $z_2$.
The total specific energy is given by $z=z_1+z_2$. Thus $\overline{z} = \overline{z_1} + \overline{z_2}$. Because the tracks originated from a beam of identical particles, and $z = \varepsilon/m$ where $\varepsilon$ and $m$ are the energy deposition and mass of the medium, we consider the same mass in the denominator of both $z_1$ and $z_2$, i.e., $m_1 = m_2$.
For two tracks with non-zero correlation and overlap (close to each other),
$\overline{z_1 z_2} \neq \overline{z_1}~\overline{z_2}$ and
\begin{eqnarray}
z_D = \frac{(\overline{z^2_1} - \overline{z_1}^2) + (\overline{z^2_2} - \overline{z_2}^2) + 2(\overline{z_1 z_2} - \overline{z_1}~\overline{z_2})}{\overline{z_1} + \overline{z_2}}
\end{eqnarray}
Therefore
\begin{eqnarray}
\Delta &=& \mu z_D =
\mu \frac{\overline{z_1}\Delta_1/\mu + \overline{z_2}\Delta_2/\mu + 2(\overline{z_1 z_2} - \overline{z_1}~\overline{z_2})}{\overline{z_1} + \overline{z_2}}\label{eq_zd_t} \nonumber \\
&=&
\frac{\Delta_1+\Delta_2}{2} + 2\mu \frac{(\overline{z_1 z_2} - \overline{z_1}~\overline{z_2})}{\overline{z_1} + \overline{z_2}} \nonumber \\
&=&
\frac{\Delta_1+\Delta_2}{2} + 2\mu \frac{{\rm cov}(z_1, z_2)}{\overline{z_1} + \overline{z_2}}
\end{eqnarray}
where we used $\overline{z_1} = \overline{z_2}$ for the first term in RHS.
By definition, the covariance of two random variables is given by
cov$(z_1, z_2) = \langle (z_1 - \langle z_1 \rangle) (z_2 - \langle z_2 \rangle)\rangle = (\overline{z_1 z_2} - \overline{z_1}~\overline{z_2})$.
In a special case of two independent $z_1$ and $z_2$, the covariance will be identical to zero. Thus, the mean \ce{^{.}OH}-radicals population is the algebraic mean of the \ce{^{.}OH}-radicals populations of two tracks, $\Delta = (\Delta_1 + \Delta_2)/2$.
The factor of 2 in the denominator is the number of tracks, which has automatically appeared. It guarantees $\Delta$ to be a track-averaged parameter, i.e., total \ce{^{.}OH}-radicals divided by the number of tracks.
In another extreme limit of two identical tracks, 1 and 2 (with a coincidental source location), $\Delta_1 = \Delta_2$, and $\Delta = \Delta_1 + \Delta_2$.
In this situation, two tracks are the constituent of a single track of a fictitious composite particle with double the charge (neglecting the Coulomb repulsion between two particles), hence there is no division by a factor of two.
Thus $(\Delta_1 + \Delta_2)/2 \leq \Delta \leq \Delta_1 + \Delta_2$, within an interval of two extreme configurations of independent and infinitely highly correlated tracks.
The latter may be realized in an idealistic and hypothetically infinitely high dose rate from a single source point.

From here one can extract formally a two-point correlation function
\begin{eqnarray}
{\rm corr}(z_1, z_2) = \frac{{\rm cov}(z_1, z_2)}{\sigma_{z_1} \sigma_{z_2}},
\end{eqnarray}
where
\begin{eqnarray}
2\mu \frac{{\rm cov}(z_1, z_2)}{\overline{z_1} + \overline{z_2}} = \Delta - \frac{\Delta_1+\Delta_2}{2} \geq 0.
\label{eq_cov}
\end{eqnarray}
Here $\sigma_{z_1}$ and $\sigma_{z_2}$ are the standard deviations of $z_1$ and $z_2$.
Eq. (\ref{eq_cov}) leads us to renormalize $\Delta$ and replace it in the rate Eqs. (\ref{Eq_rate_2}) and (\ref{Eq_rate_3}).
Formally, $\Delta \rightarrow \tilde{\Delta} = \Delta - \overline{\Delta}$ where $\overline{\Delta}$ is averaging over geometrical distribution of the tracks. Clearly, $\tilde{\Delta} \geq 0$.
For the independent tracks, with zero overlaps, $\tilde{\Delta} = 0$. Thus $\tilde{\Delta}$ is an increasing function of the overlap among the tracks.
Because $\tilde{\Delta}$ is a positive number and starts from zero (and not from a finite non-zero value), it is more convenient we reformulate our systems of equations, by replacing $\tilde{\Delta}$ for $\Delta$, starting from Eq. (\ref{eq_Qn}).
Note that the difference between $\tilde{\Delta}$ and $\Delta$ is just a shift by a constant.
By substituting $\tilde{\Delta}$ for $\Delta$ in the rate equations, we have removed all portions of intra-track recombination (track self-interactions) from our theory. 
This renormalization of the compactness parameter, $\Delta$, in the rate equations, is an analog of removal particle self-energies in many body systems and quantum field theories [\onlinecite{FetterWalecka,ItzyksonZuber}].

\subsection{H$_2$O$_2$ yield}
To calculate the rate of \ce{^{.}OH}-radicals and conversion to H$_2$O$_2$, we multiply Eq. (\ref{eq4}) by $\tilde{\Delta}$ and obtain the following non-linear equation for $\overline{n}$
\begin{eqnarray}
\frac{d\overline{n}(t)}{dt} &=& \mu \dot{z}(t) - \lambda \overline{n} - \gamma\overline{n}^2 - \eta\overline{n}^3.
\label{eq6}
\end{eqnarray}
where
$\lambda$, $\gamma$, and $\eta$ are polynomials of $\tilde{\Delta}$.
In Eq. (\ref{eq6}) $\overline{n}^2$ and $\overline{n}^3 = \overline{n}\times \overline{n}^2$ represent the population of H$_2$O$_2$ and \ce{^{.}OH}-radicals coupling with H$_2$O$_2$, respectively.
The latter represents the hydroperoxyl radical production, the scavenging reaction of \ce{^{.}OH}-radicals via \ce{^{.}OH} $+ {\rm H_2O_2} \rightarrow$ \ce{^{.}HO}$_2 + {\rm H_2O}$. 
For ease of notation, we removed the subscript, eff, from  $\lambda$, $\gamma$, and $\eta$.

For a constant dose rate, Eq. (\ref{eq6}) is integrable analytically. For a non-constant dose rate but constant dose, we split this equation into linear ($\overline{n}_0$) and non-linear ($\overline{n}_1$) terms. We calculate the solution of the linear equation using retarded Green's function method in which $\overline{n}(t)$ can be expressed in terms of an integral equation and a functional of $\dot{z}(t)$
\begin{eqnarray}
\overline{n}_0[\dot{z}(t)]
= \mu \int_{-\infty}^{t} dt' e^{-\lambda (t-t')} \dot{z}(t'),
\label{eq7}
\end{eqnarray}
the linear term in $z$.
The population yield of H$_2$O$_2$ can be calculated by
\begin{eqnarray}
Y_0[\dot{z}(t)] = \int_{-\infty}^{t} dt' \overline{n}_0^2(t')
= \frac{\mu^2}{2\lambda} G[\dot{z}(t)] z^2,
\label{eq8}
\end{eqnarray}
where
\begin{eqnarray}
G[\dot{z}(t)] = \frac{2}{z^2} \int_{-\infty}^{\infty} dt \dot{z}(t) \int_{-\infty}^{t} dt' e^{-\lambda(t-t')} \dot{z}(t').
\label{eq004703_10x} 
\end{eqnarray}
is the dose-protraction factor (see for example [\onlinecite{Sachs1997:IJRB}]).
Eqs. (\ref{eq7}) and (\ref{eq8}) constitute a linear-quadratic model for \ce{^{.}OH}-radicals pairing or H$_2$O$_2$ formation.
Eq. (\ref{eq004703_10x}) presents a cross-contribution from a two-track process that a pair of \ce{^{.}OH}-radicals (created from two individual tracks) yield in the formation of a single H$_2$O$_2$.
Note that because in this model $\lambda$ is a polynomial of $\Delta$ which itself an inter-track dependent parameter, that in turn may be affected under a change in dose and dose rate, $Y_0$, the population of H$_2$O$_2$ may not scale like $z^2$.
It may show some deviations from a quadratic power law in dose (we have seen it in our MC).

The contributions from the non-linear solutions can be calculated perturbatively in terms of a series of $\overline{n}_0$.
Considering only contribution from $\overline{n}_0$, in Sec. \ref{Sec_result_analytical_model}, we numerically calculate the dependence of $Y_0$ on the dose rates subjected to a constant dose constraint.

\begin{figure}[t]
\begin{center}
\includegraphics[width=1.0\linewidth]{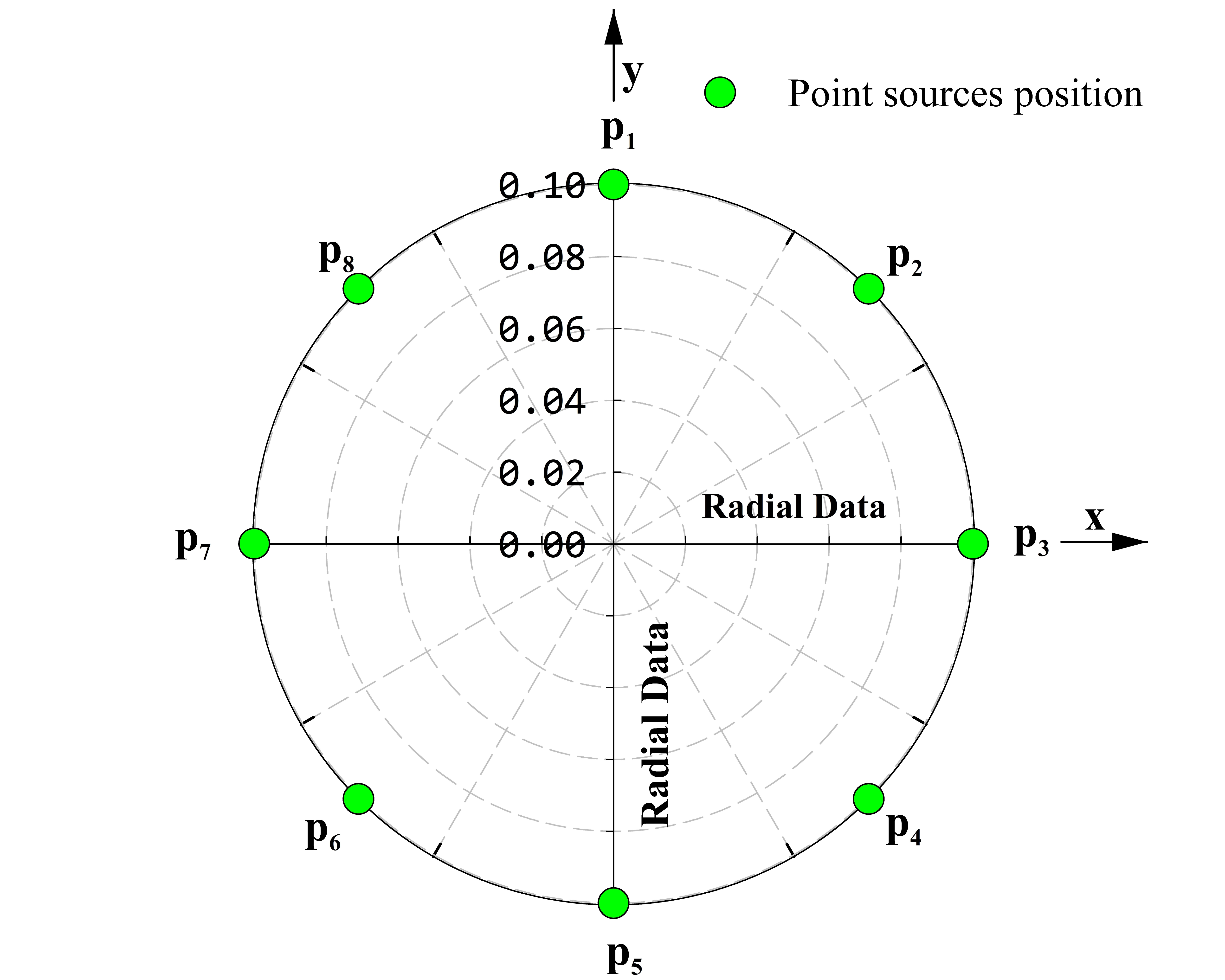}\\ 
\noindent
\caption{Schematic geometry of source of eight protons (green dots) on an $xy$ planar,
in units of nm.
}
\label{fig1}
\end{center}\vspace{-0.5cm}
\end{figure}

\subsection{Monte Carlo simulation}
\label{SubSec_Methos_MC}
To verify our analytical derivation for the population yield of \ce{^{.}OH}-radicals and H$_2$O$_2$, we perform a series of computational experiments by running chemical and UHDR modules of Geant4-DNA MC simulation of the track structure.
Many low-energy protons were initialized to impinge a cube-shaped target volume made of water with 19\% oxygen content. The physics modules \textit{G4EmDNAPhysics} (option 2) were used to simulate the physical interactions of the protons with the target.
The \textit{EmDNAChemistry} constructors are used to perform step-by-step production, diffusion, and chemical reactions of the resulting species through the end of the chemical stage, from 1 ps to 1 ms.
The thorough spectrum of species and their reactions included in the \textit{Geant4-DNA} chemistry module were used in the simulation, however, only $\ce{^{.}OH}$ and H$_2$O$_2$ populations as a function of time were selected for the analysis of the current study.
We postpone to submit our entire analysis to our forthcoming publications.

The particle source was modified into a circular $xy$-plane which was located across one side of the water phantom at $z=-16$ nm. The phantom was located at $z=0$.
The range of proton energies was chosen up to 1 MeV.
Fig. \ref{fig1} shows the schematic geometry of the protons at the source.
The phantom is a homogeneous cube of water with dimensions of 3.2 × 3.2 × 3.2 $\mu$m$^3$.
A smaller cube was considered with dimensions of 1.6 × 1.6 × 1.6 $\mu$m$^3$ to score chemical species.

As shown in Fig. \ref{fig1}, each MC simulation consists of mono-energetic protons located on a ring with a specific radius. The protons were ejected along the z-axis toward the water phantom.
We keep a fixed radius of the ring in a given simulation and score $\ce{^{.}OH}$ and H$_2$O$_2$ in water.
After performing a series of simulations we plot the population of $\ce{^{.}OH}$ and H$_2$O$_2$ vs. time and radius of the ring.

This series of simulations must be considered a computational experiment to verify our analytical predictions, resembling hypothetically the inter-track correlations of the dose rate, and examining the influence of the charged particle separation on the chemical end point of this study which is the production of the H$_2$O$_2$.

\begin{figure}
\begin{center}
\includegraphics[width=1.0\linewidth]{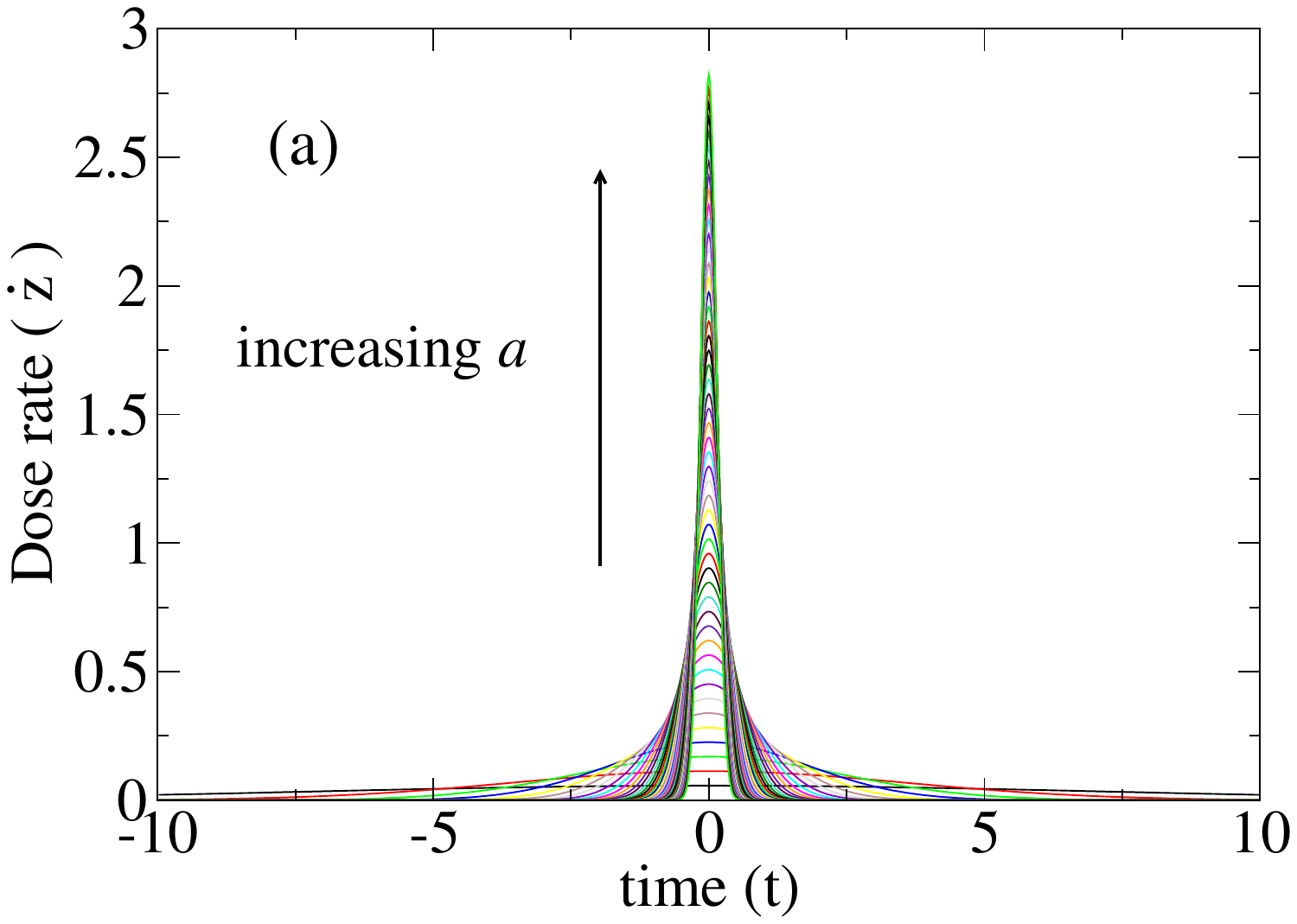} \vspace{-0.75cm} \\
\includegraphics[width=1.0\linewidth]{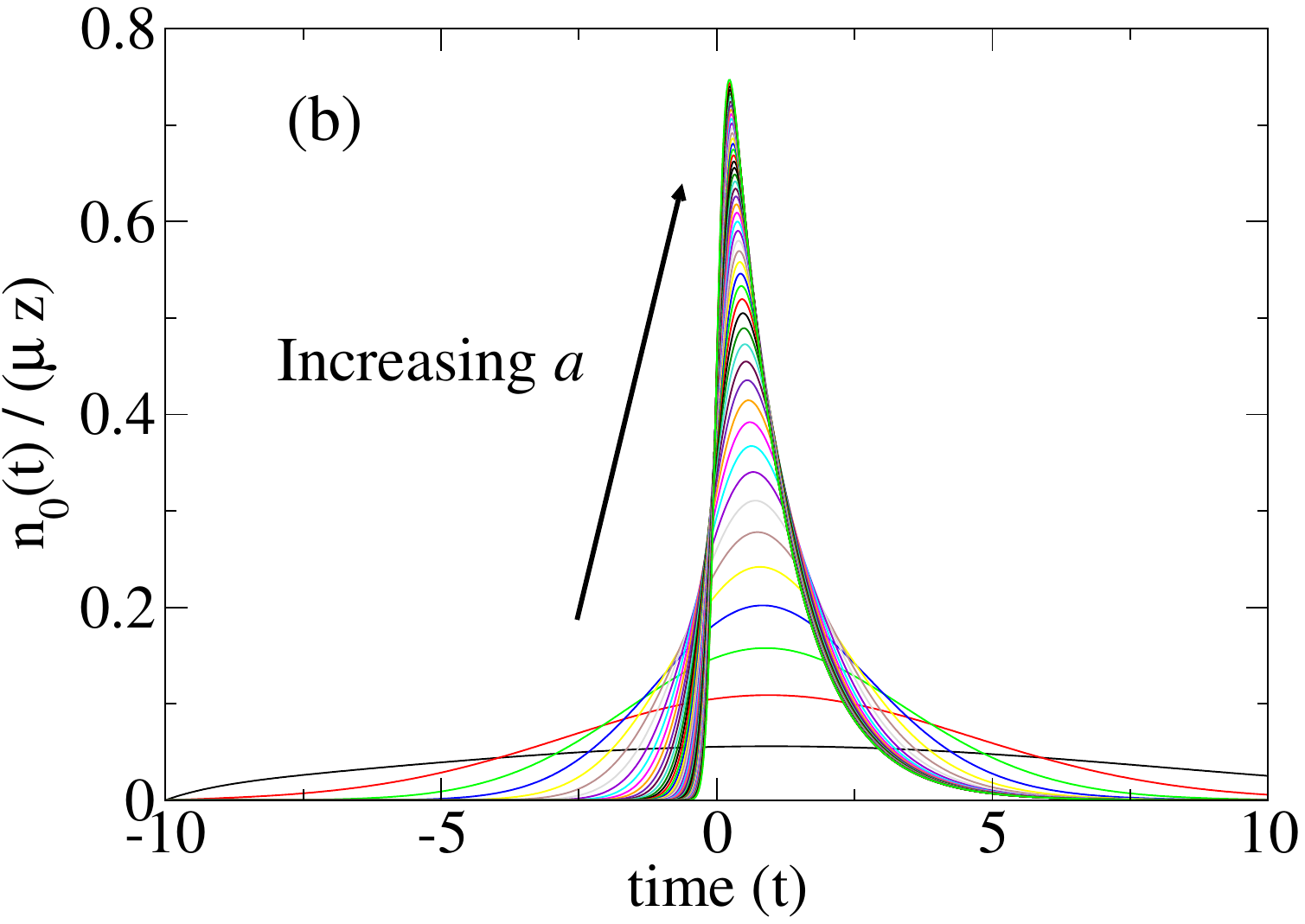} \vspace{-0.75cm} \\
\noindent
\caption{
Series of (a) instantaneous dose rates, $\dot{z}(t)$, and (b) the accumulated \ce{^{.}OH}-radicals, $n_0(t)$ as a function of time.
$a$ is a parameter that controls the dose rate.
The higher $a$, the higher the dose rate subjected to constant dose.
All areas under the dose rate curves in (a) are constant and equal to a given absorbed dose.
}
\label{fig2}
\end{center}\vspace{-0.5cm}
\end{figure}

\begin{figure}
\begin{center}
\includegraphics[width=1.0\linewidth]{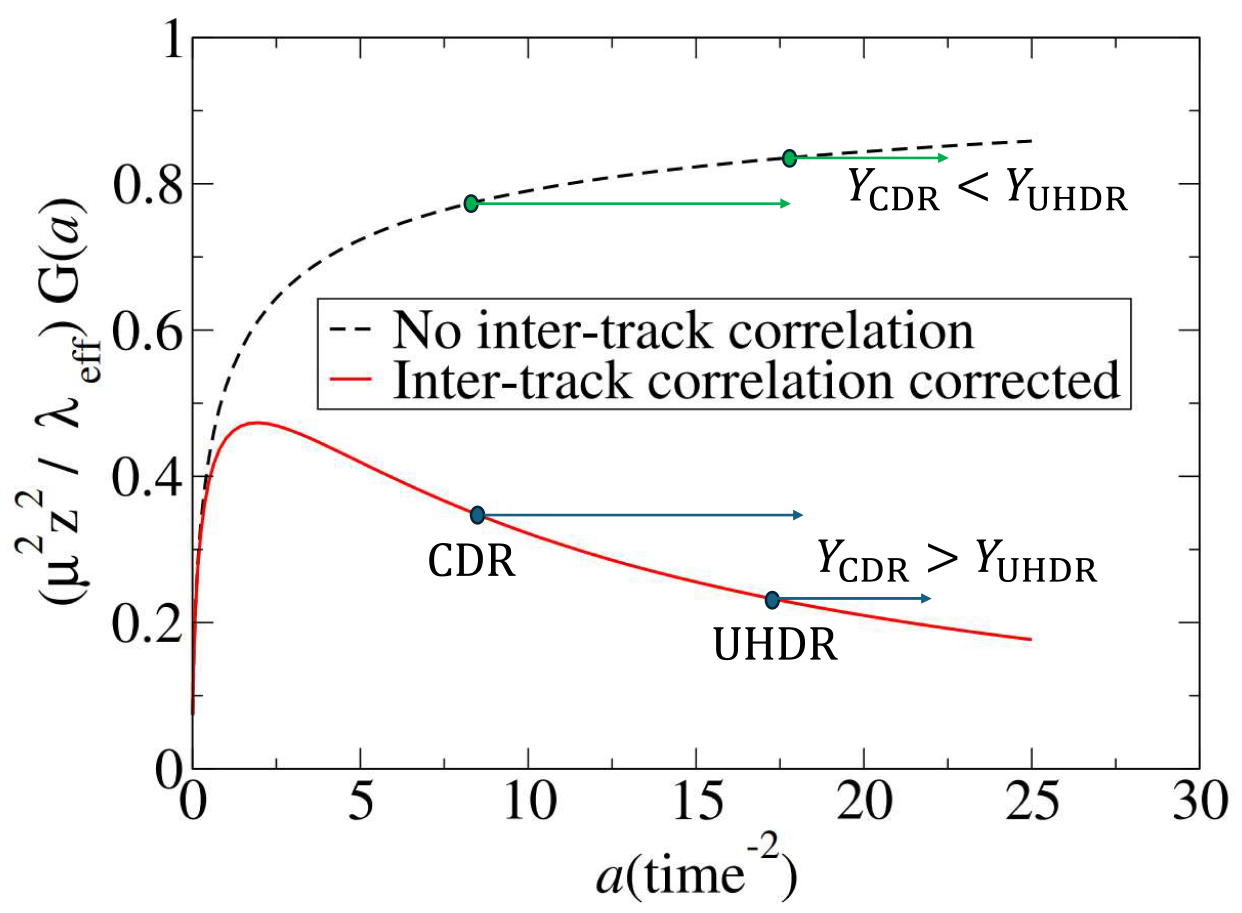} \vspace{-0.75cm} \\
\noindent
\caption{
The yield in H$_2$O$_2$ is calculated by Eq.(\ref{eq8}). The black dashed lines and the red bold line show the result of the calculation without and with inter-track correlations.
They are representative classes of solutions in the Poissonian 
($\tilde{\Delta} = 0$)
and non-Poissonian models 
($\tilde{\Delta} > 0$)
, respectively.
Two points of CDR and UHDR are given to illustrate the effect of inter-track correlations in our model for two representative points, measured and reported by Zhang et al [\onlinecite{Zhang2024:MP}]. 
}
\label{fig3}
\end{center}\vspace{-0.5cm}
\end{figure}

\section{Results}
\label{Sec_Results}

\subsection{Continuous varying dose rate - analytical model}
\label{Sec_result_analytical_model}
We illustrate the results from our analytical model for the dose rate, parameterized in a Gaussian function
\begin{eqnarray}
\dot{z}(t) = z \sqrt{\frac{a}{\pi}} e^{-a t^2}.
\label{eqGdr}
\end{eqnarray}
subjected to a constant dose constraint, $z = \int_{-\infty}^{\infty} dt \dot{z}(t) = $ const.
For a given dose and dose rate, the parameter $a$ is given uniquely. For example, 
at a dose of 10 Gy, the parameter $a$ in Eq. (\ref{eqGdr}) varies in the range of $10^{-4}/s^2$ to $100/s^2$ corresponding to the CDR and FLASH-UHDR of 0.1 Gy/s and 100 Gy/s, respectively.   
As shown in Fig. \ref{fig2}, we simulate an increase in the instantaneous dose rate, $\dot{z}(t)$, by increasing the parameter, $a$, continuously and calculate the yield in H$_2$O$_2$ as a function of dose rate, as shown in Fig. \ref{fig3}.
Note that in Fig. \ref{fig2}(b) the number of \ce{^{.}OH}-radicals that are proportional to the areas under the curves, are constant, as the variation over the dose rate was subjected to a constant absorbed dose, $z$, as a mathematical constraint.

In the limit of $a \rightarrow \infty$, i.e., at UHDR, $\sqrt{a/\pi} e^{-a t^2}$ becomes highly sharp around $t=0$. The dose rate becomes singular but integrable,
$\sqrt{a/\pi} e^{-a t^2} \rightarrow \delta(t)$.
In this limit, $G[\dot{z}(t)] = 1$, thus
\begin{eqnarray}
Y_0[\dot{z}(t)] = \frac{\mu^2}{2\lambda} z^2.
\label{eq004702_28x1}
\end{eqnarray}
For a finite $a$
\begin{eqnarray}
n_0(t) = \mu z \sqrt{\frac{a}{\pi}} e^{-\lambda t} \int_{-\infty}^{\infty} dt' e^{\lambda t'} e^{-a t'^2} \theta(t - t'),
\label{eqGdr_n0}
\end{eqnarray}
can be calculated numerically by integration over time.
In Eq. (\ref{eqGdr_n0}), $\theta(x)$ is the Heaviside step function, $\theta(x)=1$ for $x\geq 0$, and zero, otherwise. 

Fig.~\ref{fig3} shows the dependence of the H$_2$O$_2$ on $a$, which is an increasing variable proportional to the dose rate, in the Gaussian model of $\dot{z}$, given by Eq. (\ref{eqGdr}).
Thus in Fig.~\ref{fig3}, the positive direction of the horizontal axis is the direction of the increase in dose rate.
Here the population of H$_2$O$_2$, has been normalized to one (the $y$-axis), for a given $a$.
Hence, the $y$-axis in Fig.~\ref{fig3} may represent the probability in pairing \ce{^{.}OH}-radicals and forming H$_2$O$_2$ as a function of dose rate.

The black dashed curve in Fig.~\ref{fig3} corresponds to a model with no inter-track correlation, e.g., $\tilde{\Delta}=0$ or $\Delta = \overline{\Delta}$
where the only non-linear dependencies of the H$_2$O$_2$ yield on the dose rate stems from $G(a)$ as $\lambda_{\rm eff} = \gamma_1$ is a single-track parameter and dose-rate independent. Note that $\gamma_1, \gamma_2, \cdots$ in Eq. (\ref{eq1}) are the intra-track dependent parameters, only sensitive to the spatial compactness of single particle local ionization density profile and hence the particle LET.
As seen in Fig.~\ref{fig3}, the black dashed line shows a monotonic increase in the H$_2$O$_2$ yield as an increasing function of the dose rate.
It resembles one of the results of MC calculation recently presented in [\onlinecite{Derksen2023:PMB}] (see, e.g., their Fig. 4).

The red solid line corresponds to a model with inter-track correlation, $\tilde\Delta$, where
$\lambda_{\rm eff}$ is a polynomial in $\tilde\Delta$ as given by Eq. (\ref{Eq_rate_2}).
In this model, the overall H$_2$O$_2$ population reaches a maximum at an optimal value of $a$.
Beyond that, it drops continuously and 
reaches zero at $a=0$.
In Fig. \ref{fig3}, two representative points, corresponding to CDR and UHDR are selected to highlight the effect of inter-track coupling to the change of sign in H$_2$O$_2$ G-value.  
As shown, with inter-track couplings, the difference in H$_2$O$_2$ yield, $Y_{\rm CDR} - Y_{\rm UHDR}$, changes sign with and without inter-track coupling.
The positive sign, as reported by Zhang {\em et. al.} experimental data [\onlinecite{Zhang2024:MP}] strongly suggests the presence of such coupling.  
We note that performing a numerical integration to extend the results shown in Fig. \ref{fig3} to much higher dose rates requires an extension of Eq. (\ref{eq7}) that is the linear solution of Eq. (\ref{eq6}) to a non-linear model. 
As it is seen in Eq. (\ref{eq6}), in the limit of extremely high dose rates ($\mu \dot{z} >> 1$), the contribution of the non-linear terms such as $\overline{n}^2$ and $\overline{n}^3$ becomes significant.
Calculation of the numerical integration of the yield of H$_2$O$_2$ owing to the contribution of these terms in Eqs. (\ref{eq7}-\ref{eq004703_10x}) via perturbative series expansion and their domains of numerical convergence have been performed, however, because such details deviate our focus to more complex mathematical analysis beyond the scope of the present study, we postpone the presentation of the non-linear model to our future publications. 

Note that conceptually the current model is interchangeable with a similar model that describes the formation of DNA double-strand breaks by paring of single-strand breaks and the dependence of the radio-biological effects on $\Delta$ from a single track, proportional to LET [\onlinecite{Abolfath2019:EPJD}].
This similarity offers a common characteristic between the high LET and UHDR, by a single parameter, $\Delta$.

In the following section, we perform a series of MC simulations to check these ideas as sketched in Fig \ref{fig3}.


\begin{figure}[t]
\begin{center}
\includegraphics[width=1.0\linewidth]{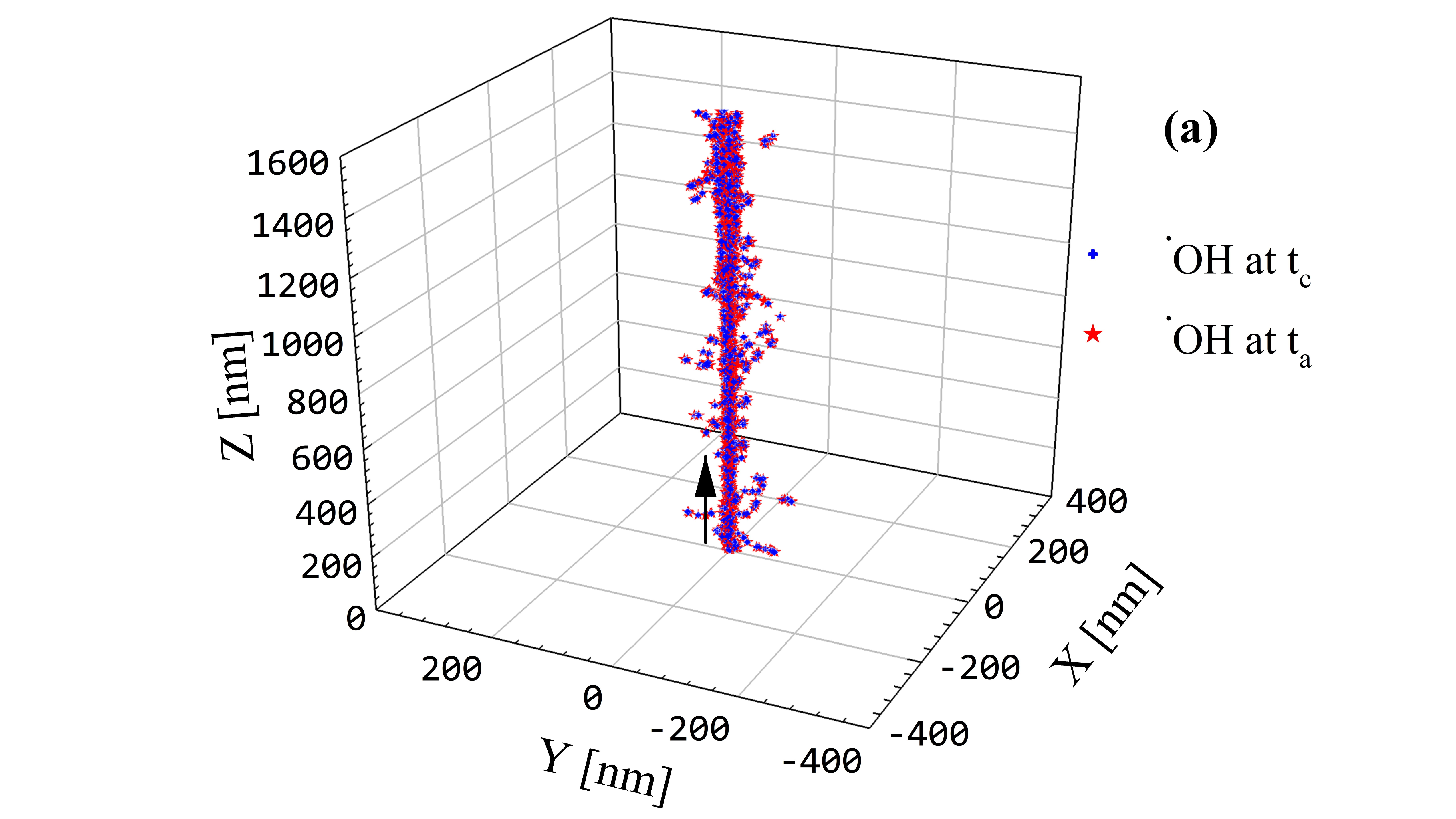}\vspace{-0.0cm} \\
\includegraphics[width=1.0\linewidth]{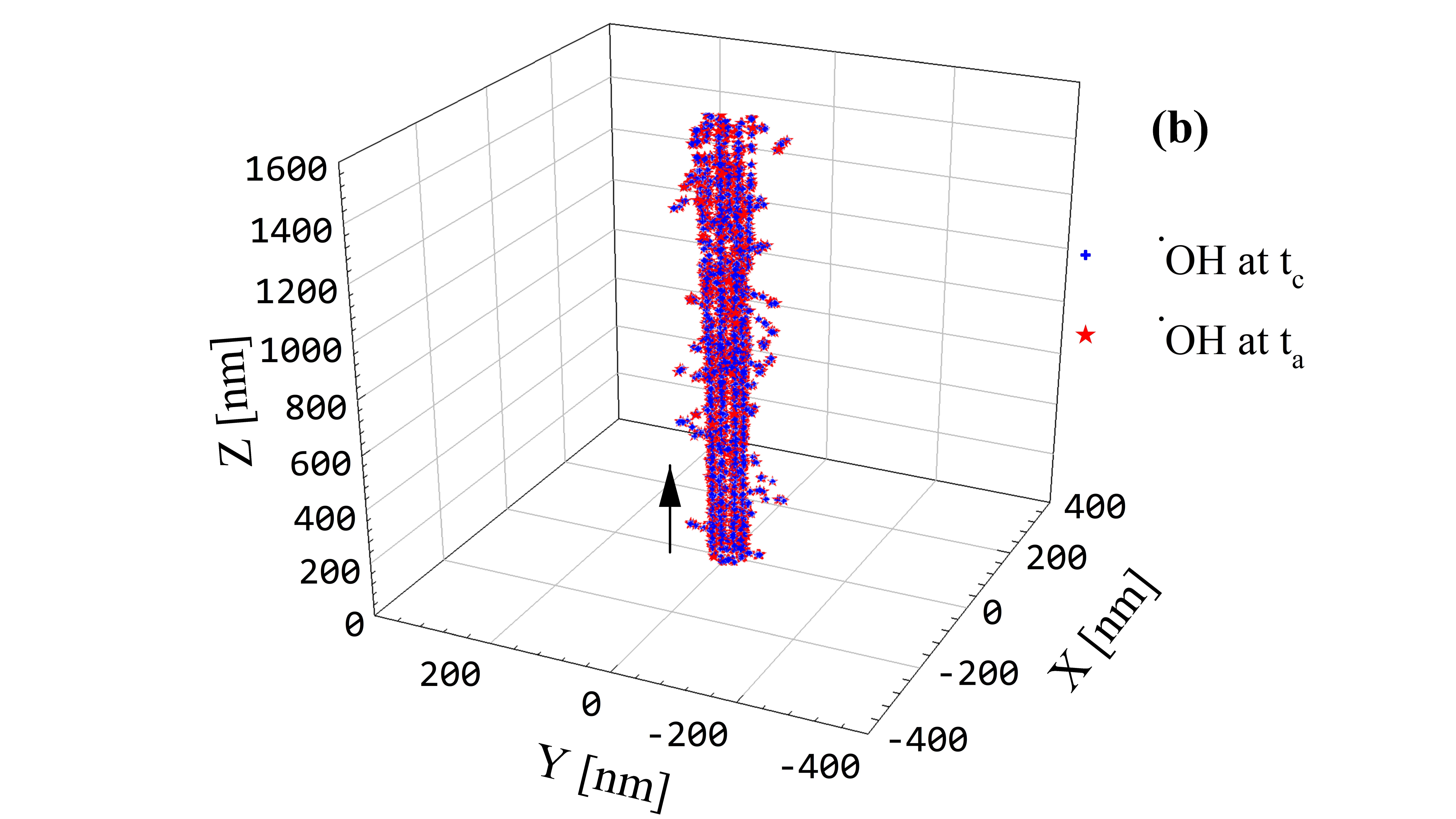}\vspace{0.0cm} \\
\includegraphics[width=1.0\linewidth]{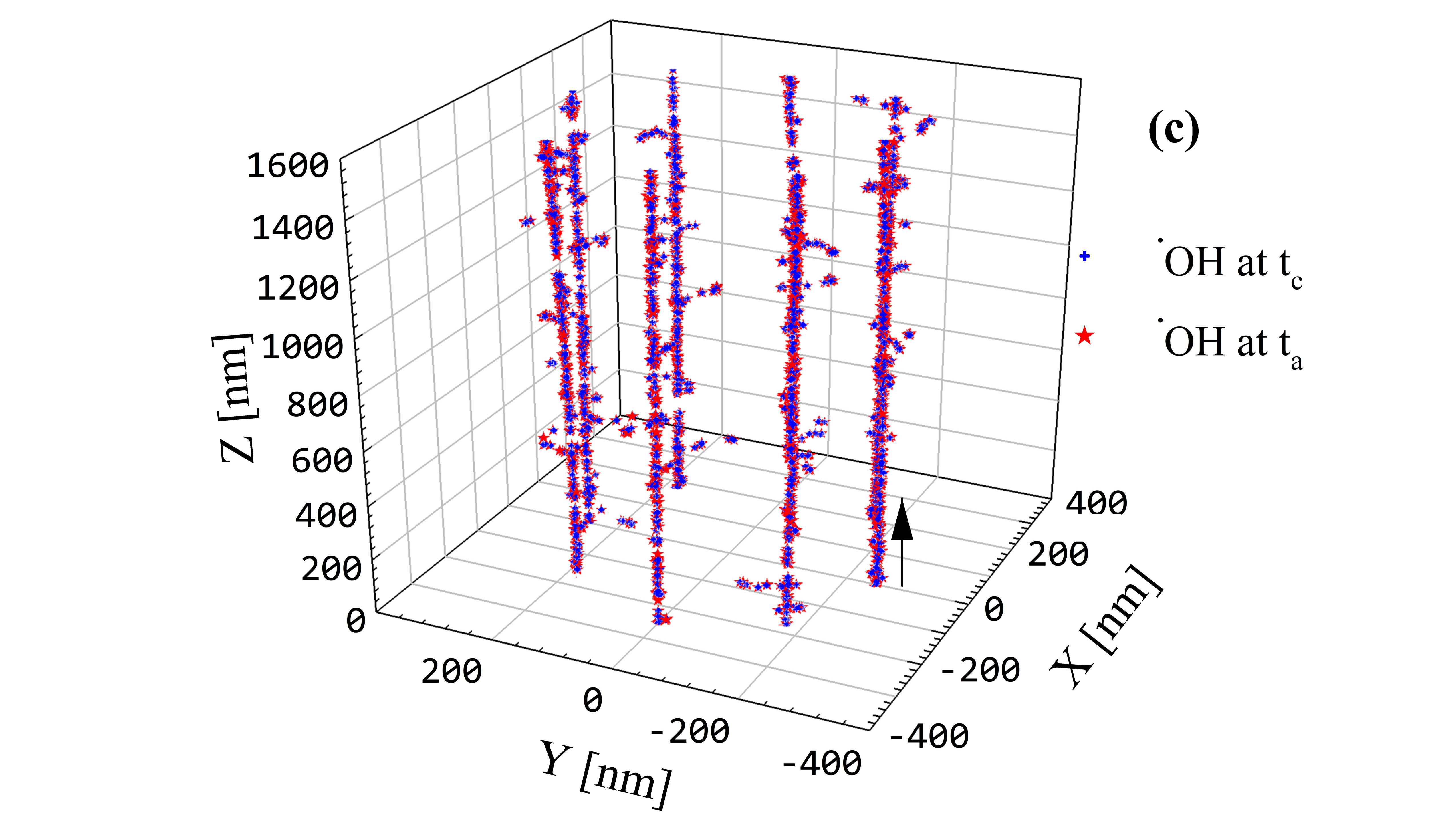}\vspace{0.0cm} \\
\noindent
\caption{
A sample of segments of eight proton tracks in a 1.6 µm length water cube (in $z$-direction), initiated from eight source points located on the planar rings with different radii. The energy of the protons at the point of entrance is 1 MeV. The blue dots show the location of \ce{^{.}OH}-radicals at the creation time t$_c$. The red dots are the location of $\ce{^{.}OH}$ after the diffusion and recombination to H$_2$O$_2$ at the annihilation time t$_a$, where a pair of $\ce{^{.}OH}$ within a certain radius, regardless to their relative orientation and magnetic polarization, combine and score a single H$_2$O$_2$.
Due to the exponential growth of the RS population, and our graphical limitations, we stopped the last frame at 10 ns.
}
\label{fig4}
\end{center}\vspace{-0.5cm}
\end{figure}

\begin{figure}
\begin{center}
\includegraphics[width=1.0\linewidth]{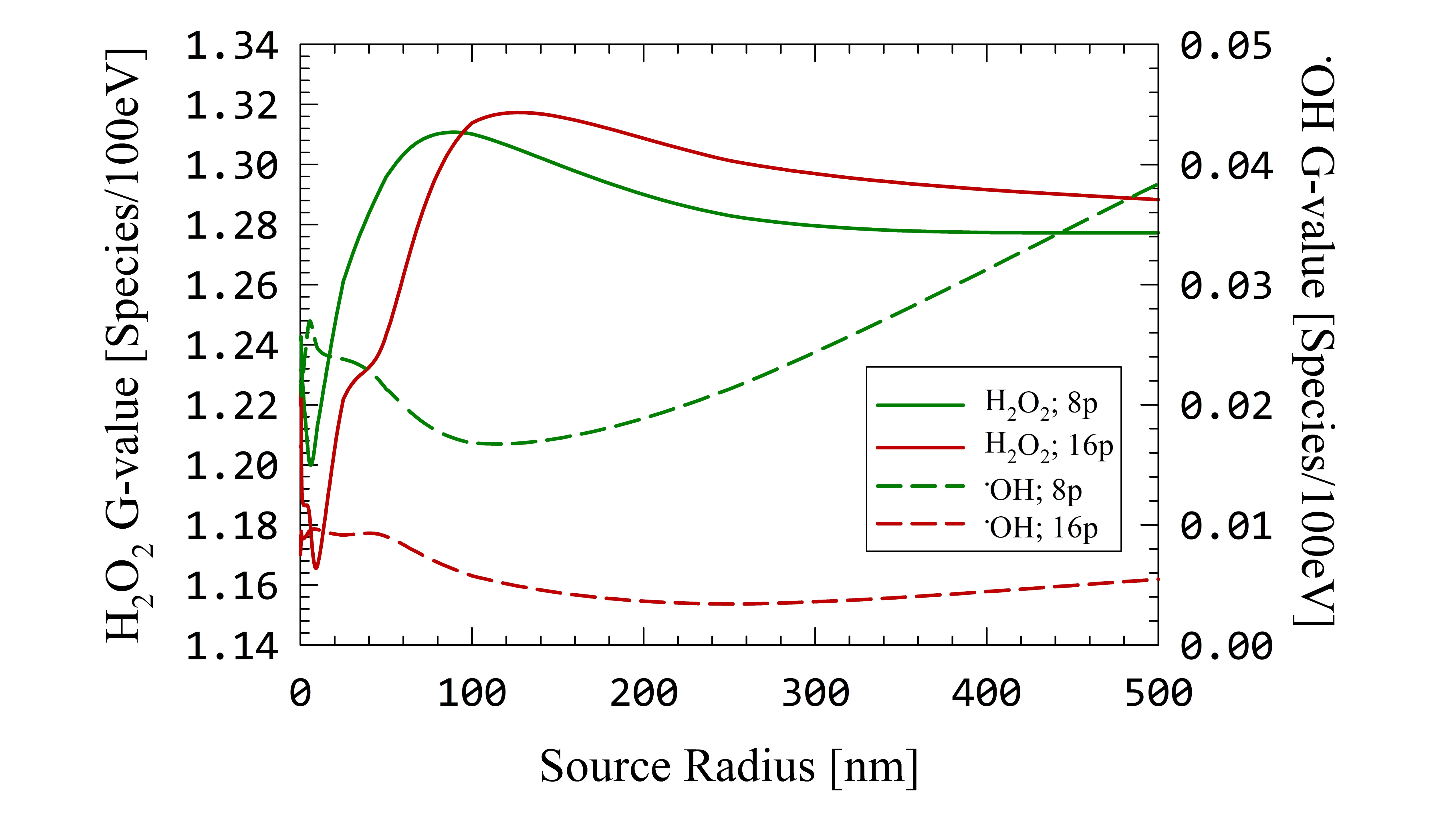} \vspace{-0.25cm}\\
\noindent
\caption{
Changes in G-value vs. the source radius of the rings were calculated for eight (green lines) and sixteen (red lines) protons with an initial energy of 1 MeV per proton in a 1.6 $\mu$m long cube.
Yield in H$_2$O$_2$ production (solid lines) and $\ce{^{.}OH}$ (dashed lines) are shown at $t=1$ms.
An increase in the H$_2$O$_2$ yield as a function of inter-track distance is an indication of the higher probabilities in $\ce{^{.}OH}$ pairing.
At a large ring radius, comparable with the diffusion length of $\ce{^{.}OH}$-radicals, the intra-track dominates the inter-track, and H$_2$O$_2$ production falls off gradually, a similar effect is shown in Fig. \ref{fig3}.
}
\label{fig5}
\end{center}\vspace{-0.5cm}
\end{figure}

\begin{figure}
\begin{center}
\includegraphics[width=1.0\linewidth]{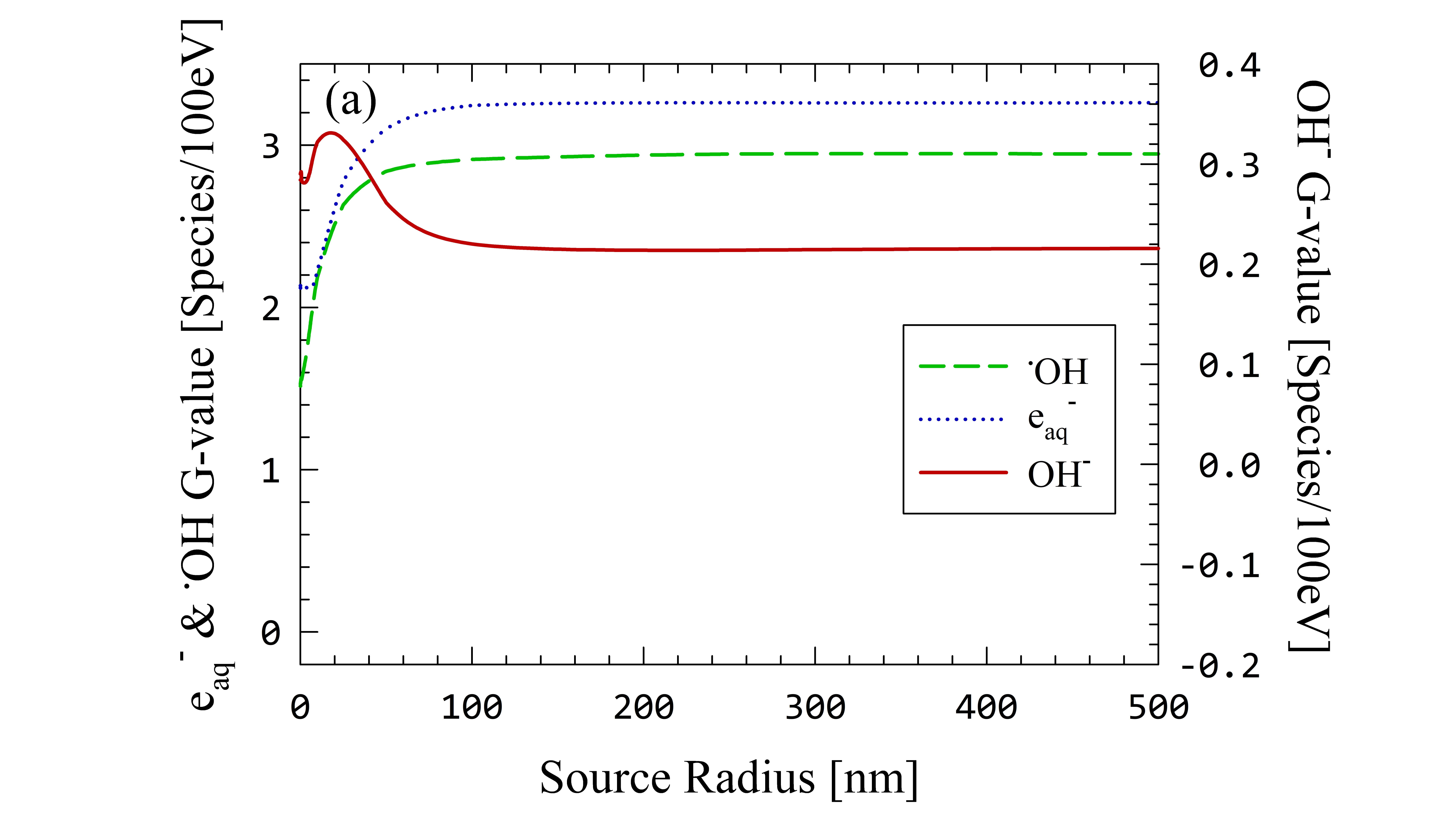} \vspace{-0.15cm}\\
\includegraphics[width=1.0\linewidth]{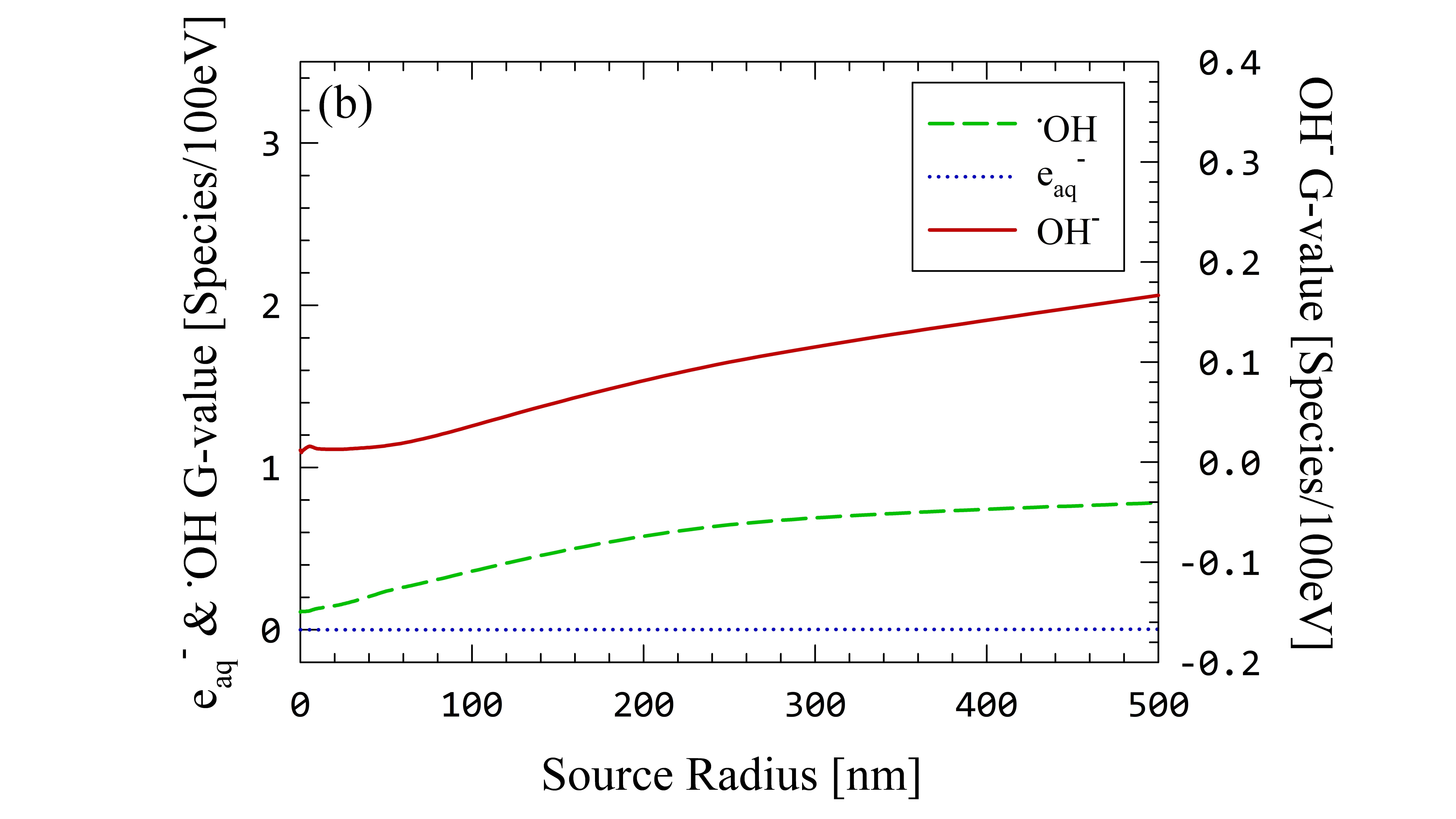} \vspace{-0.15cm}\\
\includegraphics[width=1.0\linewidth]{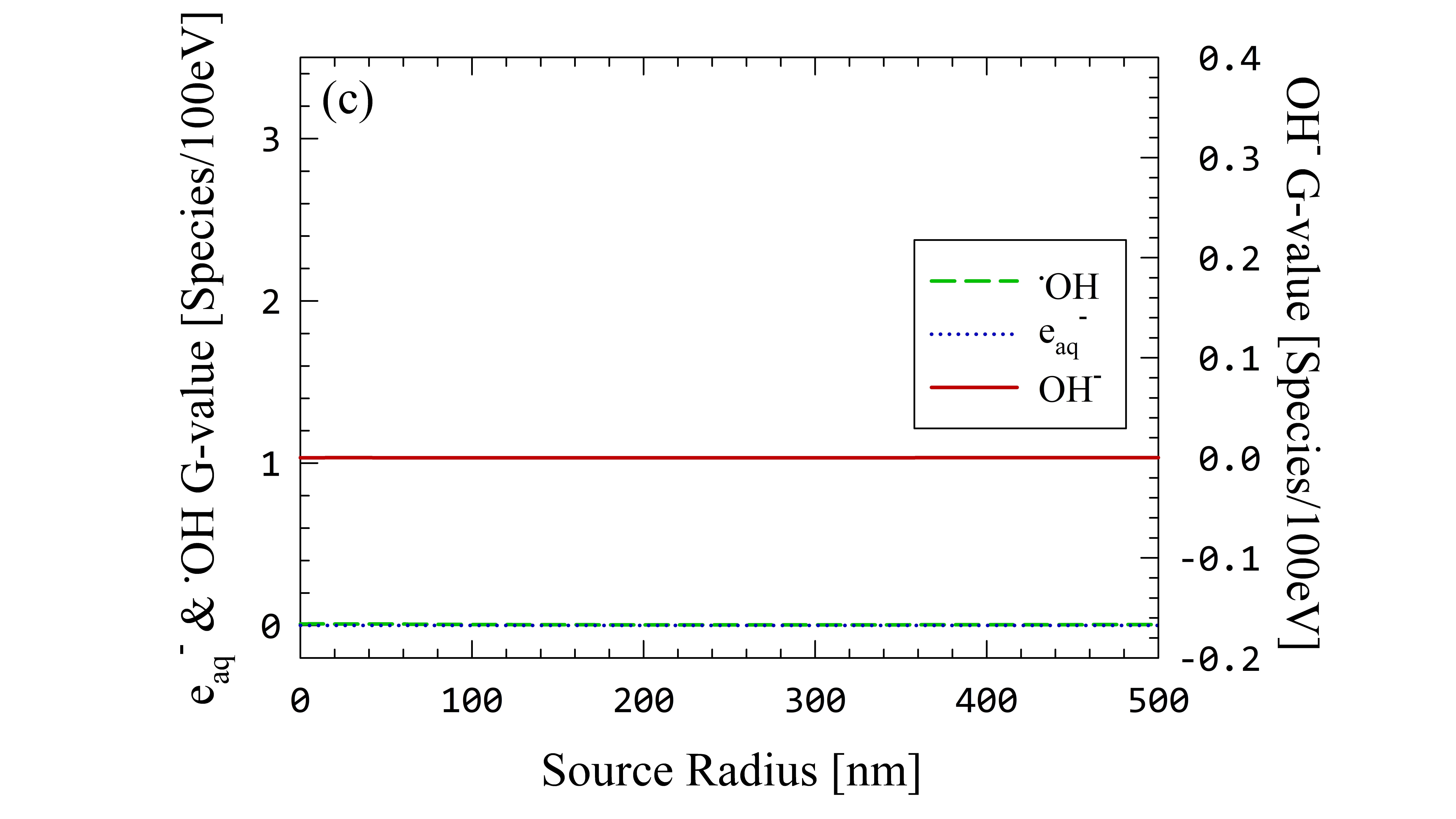} \vspace{-0.25cm}\\
\noindent
\caption{
Changes in G-value vs. the source radius of the rings were calculated for sixteen protons with an initial energy of 1 MeV per proton in a 1.6 $\mu$m long cube.
Yields in e$^-_{\rm aq}$ (blue dotted lines), $\ce{^{.}OH}$ (green dashed lines), and their product, OH$^-$ (red solid lines), at times, $t=1$ ns (a), $t=1$ $\mu$s (b), and $t=1$ ms (c) calculated by MC are shown.
The late drop in H$_2$O$_2$ yield shown in Fig. \ref{fig5} can be mainly attributed to an early peak in e$^-_{\rm aq}$-$\ce{^{.}OH}$ recombination that consumes fractions of $\ce{^{.}OH}$-radicals that could be available to form H$_2$O$_2$ in the absence of such recombination. Over this period of time, we observed negligible reactivity of $\ce{^{.}OH}$ with $\ce{^{.}H}$.
}
\label{fig6}
\end{center}\vspace{-0.5cm}
\end{figure}

\begin{figure}
\begin{center}
\includegraphics[width=1.0\linewidth]{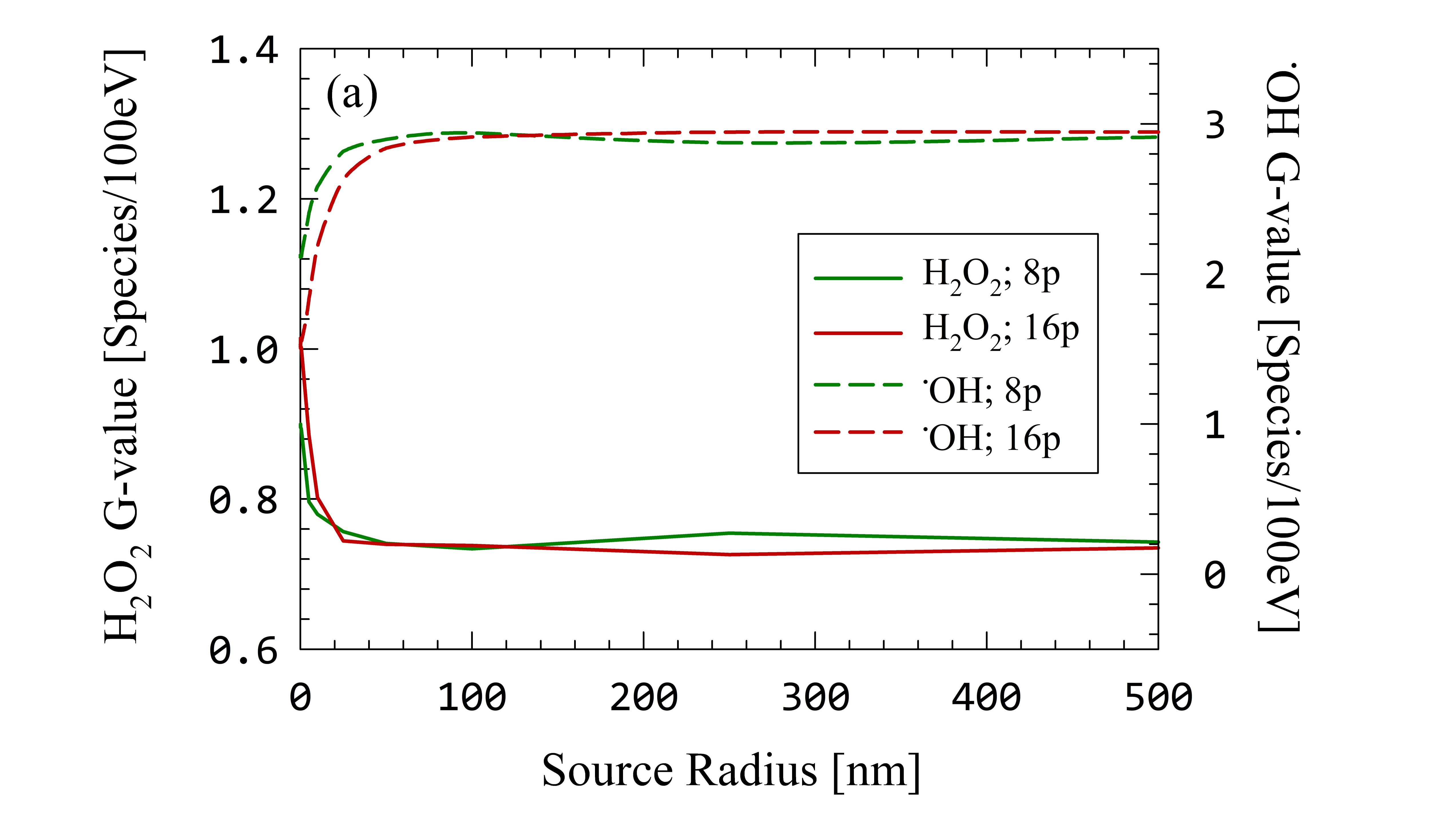} \vspace{-0.25cm}\\
\includegraphics[width=1.0\linewidth]{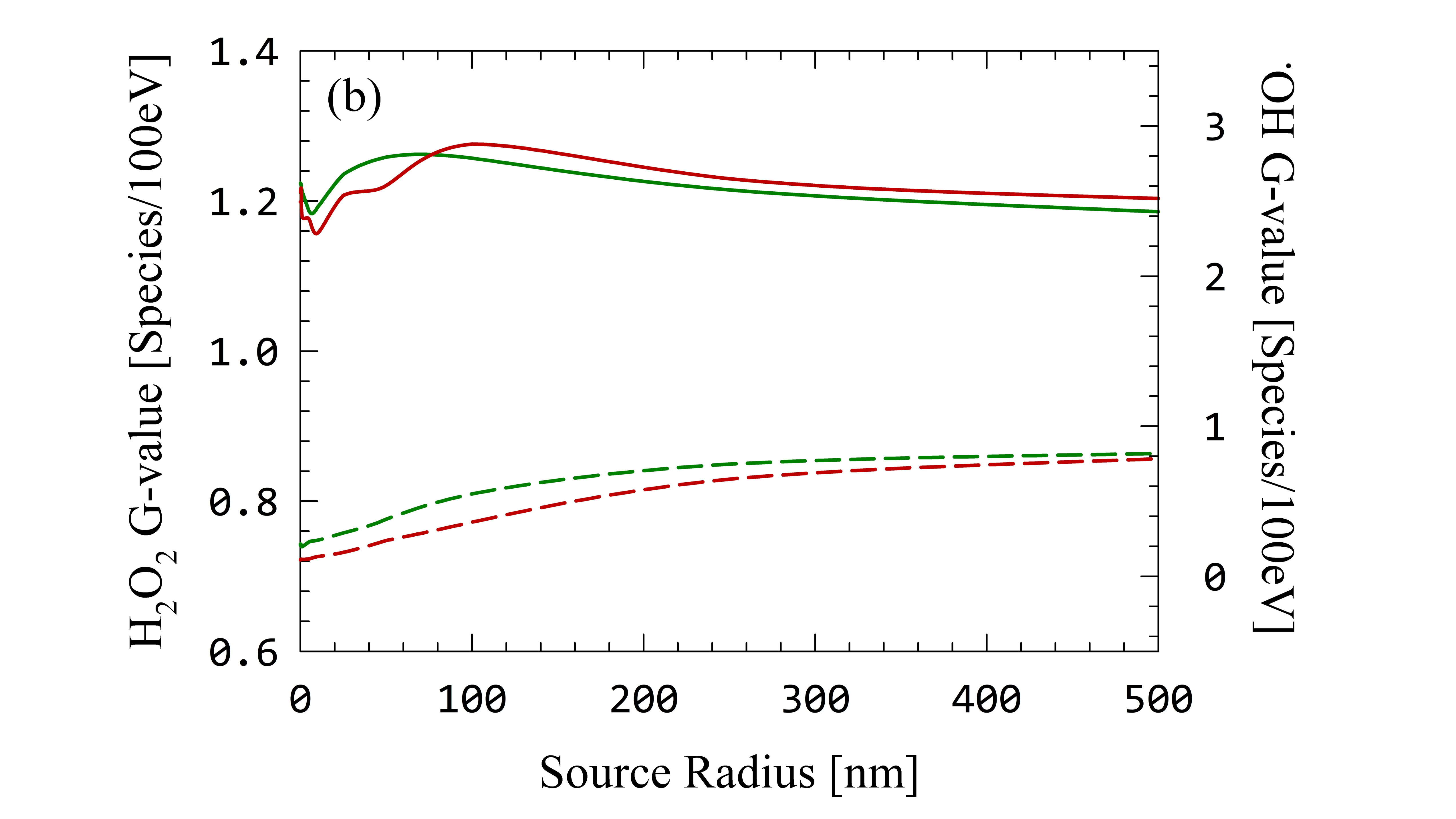} \vspace{-0.25cm}\\
\includegraphics[width=1.0\linewidth]{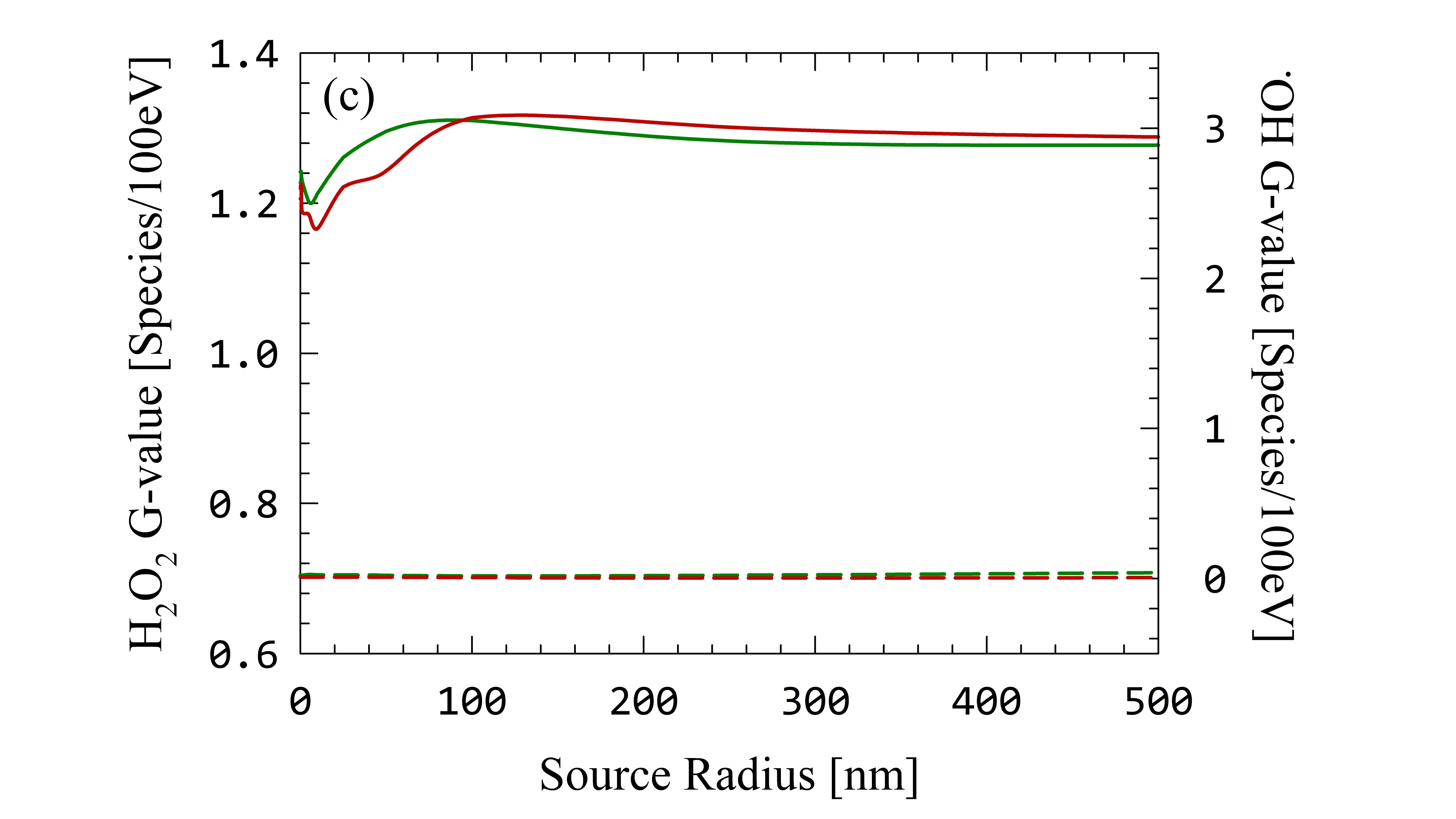} \vspace{-0.25cm}\\
\noindent
\caption{
Results are shown from the simulation of multi-track proton beam collision with a 1.6 $\mu$m long cube. Changes in G-value of species with respect to source radius demonstrated for eight (green line) and sixteen (red line) protons with entrance energy of 1.0 MeV in three times, $t=1$ ns (a), $t=1$ $\mu$s (b), and $t=1$ ms (c), for H$_2$O$_2$ and $\ce{^{.}OH}$ with values are shown in the left (H$_2$O$_2$) and the right vertical axis ($\ce{^{.}OH}$), respectively.
}
\label{fig7}
\end{center}\vspace{-0.5cm}
\end{figure}

\subsection{Monte Carlo}
\label{SubSec_G4_UHDR_results}
We employed MC simulations to investigate the effect of inter-track interaction on G-value.
Protons with 1 MeV initial energy were ejected from a ring-shaped source.
A series of ring radii, $r$, were used to fine-tune the interaction among the tracks continuously.
To have full control of the inter-track coupling, we located equally spaced protons on a series of rings with different radii.
Thus, an increase in the ring radius, $r$, is assumed to mimic a decrease in the dose rate with higher inter-track spacing.

Fig. \ref{fig4} shows the particle tracks with source radii of $r = 0.5$ nm (a), $r = 25$ nm (b), and $r = 250$ nm (c).
The largest radius demonstrates independent proton tracks, whereas the smallest one corresponds to highly correlated tracks.
In this figure, we assume the direction of the dose rate is from bottom to top, hence Fig. \ref{fig4} (a) and (c) correspond to the highest and the lowest inter-track correlations.

Fig. \ref{fig5} shows the yield in production and consumption of H$_2$O$_2$ and $\ce{^{.}OH}$-radicals stemming from eight (green lines) and sixteen (red) protons, respectively.
According to our convention, the direction of the dose rate is in the reverse direction of the $x$-axis.
We ran these MC simulations up to 1 ms with time steps of 1 ps, a one-million time-step simulation for each ring.
Within 1 ms, at a time of annihilation, $t_a$, any pair of $\ce{^{.}OH}$-radicals fell within a spherical distance smaller than the radius of 0.22 nm, were terminated and scored as a single production of H$_2$O$_2$.

The maximum yield in H$_2$O$_2$ at a characteristic ring size of $r^*$ in Fig. \ref{fig5}  
is the hallmark of these simulations.
It forms around $1\mu$s with persistent stability to 1 ms and beyond.
At first glance, it seems the maximum in H$_2$O$_2$ coincides approximately with a minimum in the population of $\ce{^{.}OH}$-radicals at 1 ms.
Such observation may give us an impression of the presence of a spatial correlation between the decay of $\ce{^{.}OH}$-radicals and the formation of H$_2$O$_2$, as both have formed at the same moment of 1ms. 
However, a more accurate inspection of coincidental chemical populations from Figs. \ref{fig5} and \ref{fig6}(a) may reveal more interesting correlations in time, rather than in space. 
It suggests the production yield of H$_2$O$_2$ at ms time scale is correlated with \ce{^{.}OH}-e$^-_{\rm aq}$ recombination and formation of OH$^-$ in ns time scale.
Similarly, below 100 nm, a ms drop in H$_2$O$_2$ population in Fig. \ref{fig7}(c) coincides with a ns peak in OH$^-$ population in Fig. \ref{fig7}(a).

The maximum of H$_2$O$_2$ (green solid line) in Fig. \ref{fig5} is located around a radius of $\approx$ 80 nm whereas the minimum of \ce{^{.}OH} (green dashed line) is located around a radius of $\approx$ 110 nm. 
Similarly, for the red lines (sixteen protons), these values are $\approx$ 120 nm and 280 nm.
Thus the spatial correlation between the populations of these species does not seem to explain the saturation effects in G(H$_2$O$_2$). 
Conversely, the maximum for G(H$_2$O$_2$) at ms seems to be correlated temporally with the drop of OH$^-$ at the same position but at a much shorter time scale, around 1 ns. 

Recall that we have interpreted the decrease in the dose rate with the decrease in the compactness of the configuration of the tracks or the inter-track recombination, parameterized by the source radius, $r$. 
The higher $r$, the lower the inter-track recombination, hence the lower dose rate. 
Therefore, an increase in G(H$_2$O$_2$) below $r = r^* \approx 100$ nm, resembles the reported experimental data that the H$_2$O$_2$ production at FLASH-UHDR is lower than CDR.

We, therefore, correlate the experimental data on the yield of H$_2$O$_2$ at UHDR with the strong inter-track coupling and the entrance of particles in closely space-packed bunches.
This explains why track structure calculations based on the uniformly spaced distribution of the tracks end up with a contradictory prediction.

As seen in Fig. \ref{fig5}, above $r^*$, the trend in H$_2$O$_2$ production reverses.
In this limit, the tracks are highly separated, and the prediction of the MC simulation gives a wrong answer to the experimental observations as frequently reported in the literature.

It is also noteworthy that by increasing the number of proton tracks from 8 and 16, we double the dose but the H$_2$O$_2$ population doesn't increase by a factor of 4 as we expected from Eqs. (\ref{eq8}) and (\ref{eq004703_10x}) assuming constant $\lambda$. Our calculation shows that $N_{H2O2}(p=16)/N_{H2O2}(p=8)$ deviates from the quadratic power law in dose as a function of separation among the tracks and the ring-source radius.

A very simplistic calculation of the diffusion length, assuming a diluted gas of $\ce{^{.}OH}$-radicals may give a ballpark for $r^*$ values.
Using the diffusion constant of $\ce{^{.}OH}$-radicals given in Geant4-DNA,
$D_f = 2.2\times 10^{-9}$ m$^2$/s, we can calculate the diffusion length using Einstein relation, $\ell=\sqrt{2D_f t}$.
Taking our maximum simulation time, $t=1$ms, we find $\ell \approx 2100$ nm.
Comparing with $r^* \approx 100$ nm in Fig. \ref{fig5}, it is clear that $\ell$ is one order of magnitude larger than $r^*$.
This is due to the high reaction rate in $\ce{^{.}OH}$-pairing to form H$_2$O$_2$, in the rate equation, Eq.(\ref{eq6}).
Moreover, our choice for the size of MC computational boxes used in this work is justifiable as they are comparable with $\ell$.
Comparing these dimensions with the dimension of the clusters, $\ell_c$, assumed for the construction of the analytical rate equation, Eq. (\ref{eq_Qn}), justifies consistency between the analytical and MC models developed in this study.  

In Geant4-DNA, $\gamma = 0.55\times 10^7$ m$^3$/(mole$\cdot$s).
Note that $\gamma$ is the rate of conversion of a pair of $\ce{^{.}OH}$-radicals to H$_2$O$_2$.
In Eq.(\ref{eq6}), there is another reaction rate, $\lambda$, that describes conversion of $\ce{^{.}OH}$-radicals to all other products (except H$_2$O$_2$).
In this context, we call $\lambda$, the scavenging rate of $\ce{^{.}OH}$-radicals.
Using the reaction rates table in Geant4-DNA, $\lambda$ can be calculated by the algebraic sum of all reaction rates (except H$_2$O$_2$).

Combining the scavenging rate in the $\ce{^{.}OH}$-radicals diffusion equation gives
\begin{eqnarray}
\overline{n}_0(\vec{r}, t) = \frac{\overline{n}_0(\vec{r}=t=0)}{4\pi D_f t} \exp\left(-\frac{r^2}{4D_f t} - \lambda t\right).
\label{Aeq34_t2}
\end{eqnarray}
From this equation, we may justify why the production of H$_2$O$_2$ reaches a maximum at a length scale much lower than the mean free path, $r^* << \ell$.
It is simply because of scavenging of $\ce{^{.}OH}$-radicals by other species originating from all tracks in the neighborhood, both intra- and inter-track scavenging events.
Diffusion is in reverse proportion to RS spatial density, i.e., the denser the volume populated by RS, the slower the diffusion.
This is the effect of molecular crowding introduced in our recent publications [\onlinecite{Abolfath2020:MP,Abolfath2023:FP}].
In this context, the term ``molecular crowding” describes the range of molecular confinement-induced effects (e.g., mobility, reactivity, long-range forces, etc.) observed in a closed system of concentrated molecules, in our case, induced in small domains by an UHDR source of radiation. 

On the other hand, the inclusion of thermal spikes boosts dramatically the diffusion constant, however, many of $\ce{^{.}OH}$-radicals may still be blocked by molecular crowding in the shock-wave wall of the thermal spikes.
Regardless of these details, it must be clear from Fig. \ref{fig5}, that a change in the distance between the particle tracks leads to a change in the G-value for $\ce{^{.}OH}$ and H$_2$O$_2$ species.

As pointed out, we may expect a significant increase in $r^*$ if the effects of thermal spikes, calculated and reported in our recent publications [\onlinecite{Abolfath2022:PMB}] were included in the diffusion constant of chemical species in MC.

For the radial distances less than 100 nm, in the current model and depicted in Fig. \ref{fig5}, the G-value for H$_2$O$_2$ shows a decreasing function of the track spacing. Conversely, $\ce{^{.}OH}$ shows an increasing trend.
The H$_2$O$_2$ yield goes up steeply until it reaches a maximum.
Beyond that track spacing, it decreases and saturates asymptotically.
These changes have the reverse trend on $\ce{^{.}OH}$.
These ascending and descending trends demonstrate the role of inter-track recombination.

The results of our calculation for the H$_2$O$_2$ G-value resemble similar behavior from a circular source with a radius of 1 nm, calculated using TOPAS n-Bio by Derksen {\em et al.} [\onlinecite{Derksen2023:PMB}], (see, Fig. 5 in that reference).
Interestingly, with an increase in radius to 100 nm, the maximum in the G-value for H$_2$O$_2$ disappears [\onlinecite{Derksen2023:PMB}] which is, indeed, in agreement with our no-correlation model, shown by black dashed lines in Fig. \ref{fig3}.

\subsection{Comparison with experimental observations}
Experimental data reported in Refs. [\onlinecite{Montay-Gruel2019:PNAS,Houda:PhDThesis,Kacem2024:SA,Zhang2024:MP}] consistently demonstrate a 10–30\% reduction in H$_2$O$_2$ yield when the dose rate increases by three orders of magnitude. This phenomenon has been observed across various radiation sources, including electrons, proton beams, and heavy particles such as carbon ions with different temporal structures, and has been corroborated using diverse experimental probes to measure hydrogen peroxide concentrations. To validate our analytical model against these experimental findings and assess its consistency with Monte Carlo (MC) simulations, we recalculated the H$_2$O$_2$ yield, using fitting parameters to qualitatively replicate the experimental trends. This approach not only confirms the model's ability to capture key features of dose-rate-dependent behavior but also strengthens its relevance in bridging experimental observations with theoretical predictions.

\begin{figure}
\begin{center}
\includegraphics[width=1.0\linewidth]{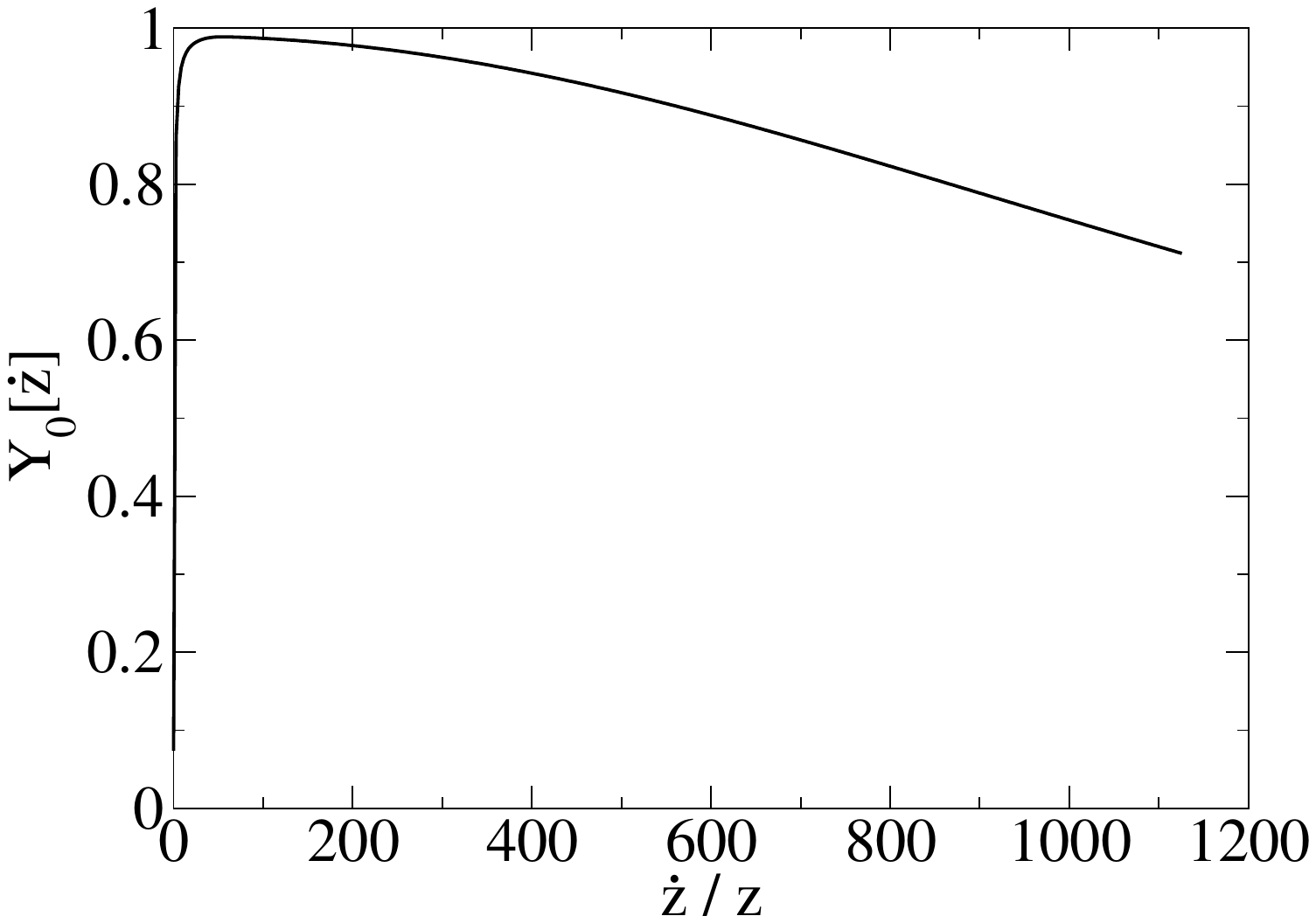} \vspace{-0.75cm} \\
\noindent
\caption{
The yield in H$_2$O$_2$, $Y_0[\dot{z}]$, as a function of $\dot{z}/z$, calculated by Eq.(\ref{eq8}) with the fitting parameters to replicate the recent experimental observations. 
With a constant dose, $z$, and three orders of magnitude larger in dose rate, $\dot{z}$, our model shows a reduction in H$_2$O$_2$ yield by approximately 20\%, in agreement with our MC results and the 
reported measurements by Zhang et al [\onlinecite{Zhang2024:MP}]. 
By adjusting the fitting parameters the percentage in reduction of $Y_0$ can go below 5-10\%. 
}
\label{fig8}
\end{center}\vspace{-0.5cm}
\end{figure}

To this end we recall Eqs. (\ref{Eq_rate_2}), and (\ref{eq8}-\ref{eq004703_10x}). 
Knowing that $\Delta = \Delta(a)$ is an analytical function of the parameter $a$, and it increases as the dose rate increases, we may expand it in terms of polynomials of $a$, with the following Taylor's series expansion, $\Delta(a) = \Delta_0 + \Delta_1 a + \frac{1}{2} \Delta_2 a^2 + \cdots$, to obtain 
\begin{eqnarray}
\lambda(a) = \gamma_1 
+ \frac{\gamma_2}{2} \Delta_1 a + \cdots.
\label{eq004702_28x1x}
\end{eqnarray}
Note that at the limit of $a=0$, $\Delta(a)=0$, hence $\Delta_0 = 0$.
Insertion of Eq. (\ref{eq004702_28x1x}) into Eqs. (\ref{eq8}-\ref{eq004703_10x}) and using the expansion coefficients as the free parameters we can fit $Y_0$ to the experimental data.
A typical result of such fitting is shown in Fig. \ref{fig8}.
Here the horizontal axis represents the dose rates at their peaks normalized by the dose where at $t=0$, the dose rates reach their maximum values as $\dot{z}/z = \sqrt{a/\pi}$ (see Fig. \ref{fig2}). 
Considering $z$, the deposited dose by a single pulse, $z/\dot{z}$ represents a single pulse irradiation time, or the pulse width.  

As illustrated in Fig. \ref{fig8}, at a given dose, $z$, with an increase in $\dot{z}$, the yield in H$_2$O$_2$ decreases slowly. 
In this example with an increase of $\dot{z}$ to three orders of magnitude, $Y_0$ drops approximately 20\%.
Conversely, at a given dose rate, $Y_0$ is an increasing function of dose, as it should be. 
This illustrative example shows how experimentalists can fit their data into the present model by using the free parameters in this model. 
The numerical values of the fitting parameters can be used to describe the distribution of the particles in the beam, thus the present model can be served as part of beam parameters characterization.

\begin{figure}
\begin{center}
\includegraphics[width=1.0\linewidth]{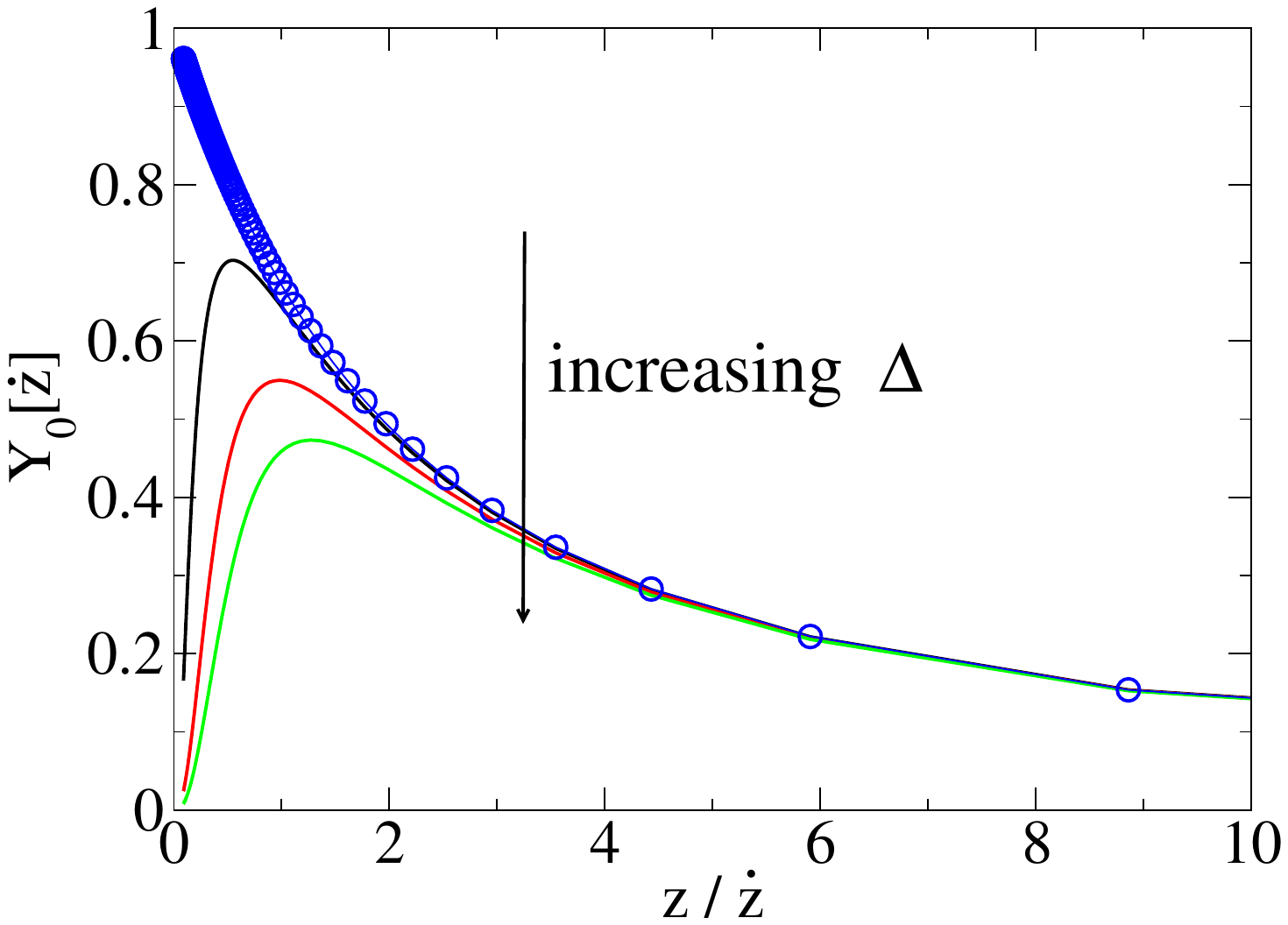} \vspace{-0.75cm} \\
\noindent
\caption{
The yield in H$_2$O$_2$, $Y_0[\dot{z}]$, as a function of $z/\dot{z}$, calculated by Eq.(\ref{eq8}). The blue circles are the points with $\Delta = 0$. 
}
\label{fig9}
\end{center}\vspace{-0.5cm}
\end{figure}

Alternatively, we plot $Y_0$ in a dual representation of Fig. \ref{fig8} as shown in Fig. \ref{fig9} where the horizontal axis is the ratio of dose and dose rate ratio, $z/\dot{z}$, or a single pulse width. Along the $x$-axis, $z$ is constant. 
Considering $\Delta$ an independent parameter (e.g., representing track-clustering index) we observe that 
by increasing $\Delta$, with a constant $z$ and $\dot{z}$, $Y_0$ suppresses from black to green curves.
The blue curve corresponds to $\Delta=0$. 


\begin{figure}
\begin{center}
\includegraphics[width=1.0\linewidth]{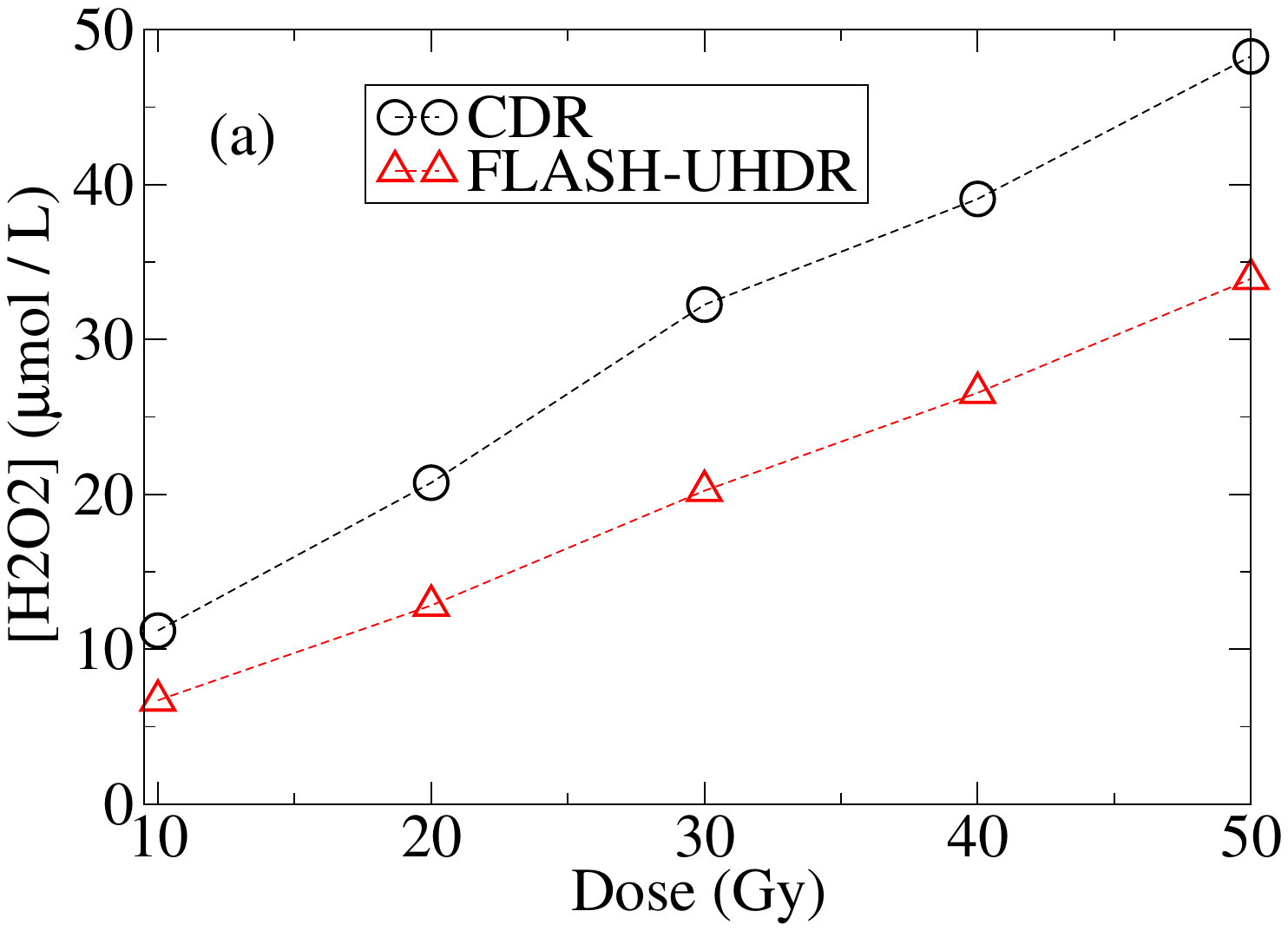} \vspace{-0.75cm} \\
\includegraphics[width=1.0\linewidth]{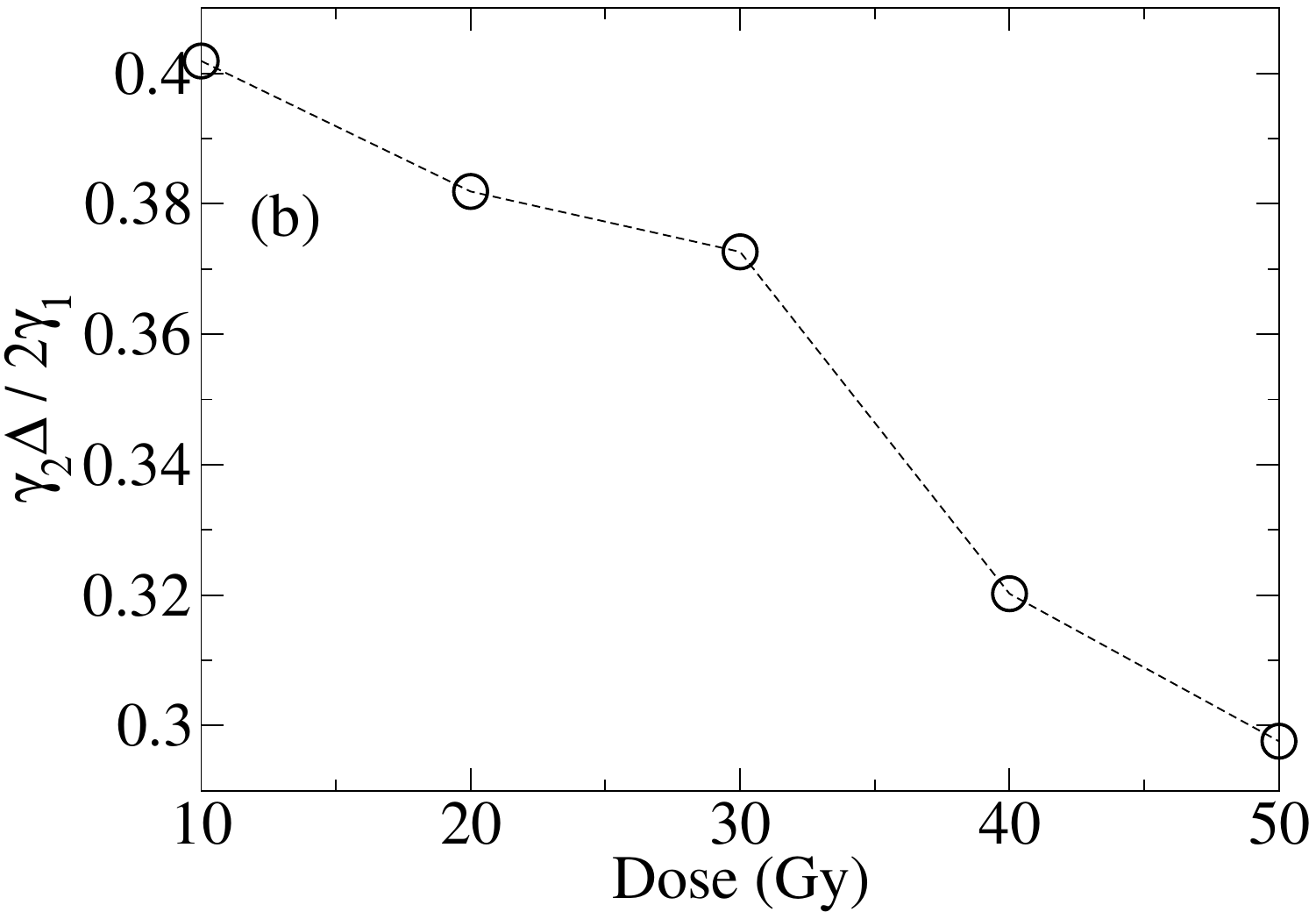} \vspace{-0.75cm} \\
\noindent
\caption{
(a) The measured yield in H$_2$O$_2$, as a function of dose and dose rate, reported experimentally in Refs. 
[\onlinecite{Houda:PhDThesis,Kacem2024:SA,Zhang2024:MP}], 
and (b) fitted to Eq. (\ref{Eq_rate_2xxx}). 
}
\label{fig10}
\end{center}\vspace{-0.5cm}
\end{figure}

The change in dose rate in the current experimental data, however, is not continuous.
The data rather contain a comparison of $Y$ for two dose rates of low and high within a range of doses used in clinical applications.
We, therefore, build up a fitting scheme based on such representation of the experimental data by interpreting $\Delta$ a chemical heterogeneity index.
Fig. \ref{fig10} shows the yield of H$_2$O$_2$ as a function of dose at two dose rates of CDR and FLASH-UHDR, 
reported in Refs. [\onlinecite{Houda:PhDThesis,Kacem2024:SA,Zhang2024:MP}].
These data were collected using an electron beam linear accelerator with energy 6 MeV (eRT6) with the operational modes at 555 Gy/s (FLASH-UHDR) and 0.1 Gy/s (CDR).
In this data, non-buffered water was irradiated at the normal level of oxygen, 21\%. The yields of hydrogen peroxide were measured minutes after the irradiation and so after the homogenous phase of chemistry where 
H$_2$O$_2$ diffused uniformly throughout the sample. We perform the fitting starting from a polynomial expansion of the yield equation, Eqs. (\ref{eq8}) and (\ref{Eq_rate_2})
where
\begin{eqnarray}
Y_{\rm eff}[\dot{z}(t)] &=& \frac{\mu^2}{2\lambda_{\rm eff}(\Delta)} G[\dot{z}(t)] \overline{z^2} \nonumber \\ &\approx&
\frac{\mu^2}{2} G[\dot{z}(t)] \overline{z^2} \left[\gamma_1+\frac{\gamma_2}{2!}\Delta + {\cal O}(\Delta^2)\right]^{-1} \nonumber \\ &\approx&
\frac{\mu^2}{2\lambda} G[\dot{z}(t)] \overline{z^2} \left[1-\frac{\gamma_2}{2\gamma_1}\Delta\right] + {\cal O}(\Delta^2),
\label{eq8x}
\end{eqnarray}
where $\overline{z^2} = (D + \Delta/\mu) D$, and $\overline{z} = D$, hence Eq. (\ref{eq8x}) describes a linear-quadratic model for H$_2$O$_2$ yield. A mathematical proof of this equation is given in the appendix. 

Using
\begin{eqnarray}
\lambda_{\rm eff}^{-1}(\Delta) &=& \left[\gamma_1+\frac{\gamma_2}{2!}\Delta+\frac{\gamma_3}{3!}\Delta^2+\frac{\gamma_4}{4!}\Delta^3 + \cdots\right]^{-1} \nonumber \\ &\approx&
\lambda^{-1} \left[1-\frac{\gamma_2}{2\gamma_1}\Delta\right] + {\cal O}(\Delta^2),
\label{Eq_rate_2x}
\end{eqnarray}
and $\lambda = \gamma_1$. 
Assigning $Y_{\rm FLASH} = Y_{\rm eff}$ and $Y_{\rm CDR} = Y$, we obtain
\begin{eqnarray}
Y_{\rm FLASH} = Y_{\rm CDR} \left[1-\frac{\gamma_2}{2\gamma_1}\Delta\right]
\label{Eq_rate_2xx}
\end{eqnarray}
hence 
\begin{eqnarray}
\frac{Y_{\rm CDR} - Y_{\rm FLASH}}{Y_{\rm CDR}} = \frac{\gamma_2}{2\gamma_1}\Delta.
\label{Eq_rate_2xxx}
\end{eqnarray}
Note that $\gamma_1$ is the rate of decay of \ce{^{.}OH}-radicals to any processes except the formation of H$_2$O$_2$, and $\gamma_2$ is the rate of decay of \ce{^{.}OH}-radicals only to the formation of H$_2$O$_2$.
For simplicity for the calculation of $\gamma_1$, we consider only the contribution of the most prominent reaction observed in our MC simulation, presented in the preceding section, \ce{^{.}OH} + $e^-_{\rm aq} \rightarrow$ OH$^-$  
corresponding to the reaction rate $\gamma_1 = 2.95\times 10^7 {\rm m}^3/({\rm mole}\cdot s)$. 
Similarly, for \ce{^{.}OH} + \ce{^{.}OH} $\rightarrow$ H$_2$O$_2$ the reaction rate is given by $\gamma_2 = 5.5\times 10^6 {\rm m}^3/({\rm mole}\cdot s)$. 
Thus we find $\gamma_2/(2\gamma_1) = (5.5\times 10^6)/(2\times2.95\times 10^7) \approx 0.1$, hence 
\begin{eqnarray}
\frac{Y_{\rm CDR} - Y_{\rm FLASH}}{Y_{\rm CDR}} \approx \frac{1}{10}\Delta.
\label{Eq_rate_2xxxx}
\end{eqnarray}
Eq. (\ref{Eq_rate_2xxxx}) serves to extract $\Delta$ from the experimental data, as illustrated in Fig. \ref{fig10}. 
Note that the input parameters in our MC calculation contain all possible reaction rates and diffusion constants of the radiolysis species available in UHDR example of Geant4-DNA. Those interactions contribute to $\gamma_1$ with the weights averaged over the kinematic of species deduced from MC due to their mobilities and thermal diffusion. Here, for the sake of simplicity, we neglected all other interactions except \ce{^{.}OH} + $e^-_{\rm aq} \rightarrow$ OH$^-$.

\section{Discussion and conclusion}
\label{Sec_Conclusion}
The relevance of inter-track interaction has recently been examined by Thompson {\em et al.}  [\onlinecite{Thompson2023:PMB}] by MC simulation of interacting proton tracks, where no significant changes in \ce{^{.}OH}-radicals or H$_2$O$_2$ yields was found at clinically relevant doses. This claim is also supported by other simplistic geometric track overlap models.

These computational results stand in contrast with the experimental evidence reported by  Blain {\em et al.} [\onlinecite{Blain2022:RR}] who measured lower H$_2$O$_2$ yields at UHDR compared to CDR. 
Furthermore, Montay-Gruel {\em et al.} [\onlinecite{Montay-Gruel2019:PNAS}]  and Kacem {\em et al.}[\onlinecite{Kacem2022:IJRB}] observed a similar decrease in H$_2$O$_2$ yields at UHDR with electrons. 
These results were repeated recently and compared with the standard model calculation revealing apparent contradictions between theory and experiment [\onlinecite{Houda:PhDThesis,Kacem2024:SA,Zhang2024:MP}].
This discrepancy, as well as the observation that MC simulations tend to measure an increase in H$_2$O$_2$ yields as dose rate increases as opposed to the experimentally measured decrease in H$_2$O$_2$ yields, suggests that the current model calculations and MC simulations do not provide an adequate representation of the interdependent chemical reactions occurring in the irradiation of oxygenated water.

In this work, we developed a model to fill the gap between theory and experiment. Our results provide systematic solutions for such discrepancies to add inhomogeneities in track distribution to resolve the controversy. The current disagreement between theory and experiment may have to do with the limitations, assumptions, and lack of inter-track correlations in the initial conditions of the user-controlled MC simulations in accounting for the clustering of the particle tracks which is a central assumption in any MC simulation.

We have demonstrated the impact of inter-track interactions at UHDR on the structural heterogeneity and the chemical reactions of RS.
The results have shown that the heterogeneous chemical phase from bunching the charged particles in a beam at UHDR can provide a logical interpretation for the experimental observations.
Whereas the current models, assume a uniform distribution of the tracks, hence the authors believe the homogeneous chemical phase is unable to predict the experimental data.

Under such inter-track initial conditions, implemented in our MC setup, we have observed that the reduction of H$_2$O$_2$ dose rate effect in ms time-scale is mainly due to the early time-scale (heterogenous phase of chemistry) of e$^-_{\rm aq}$ reactions with \ce{^{.}OH}-radicals, with a peak around ns.

This interaction is particularly significant under conditions of high inter-track coupling, as demonstrated in our simulations. For example, as inter-track distances increase, the time of peak e$^-_{\rm aq}$ - \ce{^{.}OH} reactions is delayed, resulting in lower G-values for \ce{^{.}OH} formation and reduced H$_2$O$_2$ yields. The recombination of \ce{^{.}OH} and e$^-_{\rm aq}$ contributes to these effects, with clear temporal correlations emerging across several orders of magnitude, from nanoseconds to milliseconds.

The influence of species such as e$^-_{\rm aq}$ on the decay rate, $\lambda$, in our analytical rate equation, is therefore consistent with the results of MC, shown in Fig. \ref{fig6}. 
To have a better visualization of the effect of the time evolution of 
e$^-_{\rm aq}$ reactions with \ce{^{.}OH}-radicals on the formation of H$_2$O$_2$, we plotted similar graphs in Fig. \ref{fig7} for eight and sixteen protons on the circular rings with the same time sequence as in Fig. \ref{fig6}. 
By increasing the inter-track distances, the peak in e$^-_{\rm aq}$ - \ce{^{.}OH} reaction occurs with a time delay and with a lower G-value.
More specifically, the peaks in G-value of OH$^-$, that is the product of e$^-_{\rm aq}$ - \ce{^{.}OH} reaction, for low source radii (high inter-track coupling) is above 0.3 at 1 ns.
For high source radii (low inter-track coupling), this G-value is shifted to below 0.2, with a time lag of 1 $\mu$s.
Additionally, the rate of the reduction in the G-value of OH$^-$ at high inter-track couplings is significantly higher than in low inter-track couplings.
A similar turning point visible for the drop of G(H$_2$O$_2$) between 0 and $\approx$ 15 nm can be explained by the recombination of $\ce{^{.}OH}$ and e$^-_{\rm aq}$ and formation of OH$^-$ at 1 ns.
It is an indication of temporal correlation with the turning point of G(OH$^-$) between 0 and 15nm as shown in Fig. \ref{fig6}(a), i.e., the same chain of events leading to the formation of H$_2$O$_2$ at 100 nm where the correlation in time with six orders of magnitude time lag, ns to ms sequence of reactions, resulted because of the turning point of G(OH$^-$) at 1 ns, as shown in Fig. \ref{fig6}(a).

We therefore conclude that the differences in the composition of the secondary species, with two mutually exclusive conditions of correlated vs. non-correlated tracks, and nano- vs. $\mu$-seconds peaks in
e$^-_{\rm aq}$ - \ce{^{.}OH} reaction, affect the yield and formation of H$_2$O$_2$ end-point, consistent with the reported experimental data.
We also observed that \ce{^{.}OH}-radicals do not significantly react with \ce{^{.}H} under these circumstances.

As seen in Fig. \ref{fig7}, by increasing the number of protons from eight to sixteen, the difference between the yields between the largest and lowest source radii increases. 
With an increase in the number of tracks, hence the track compactness ($\Delta$ in our analytical model), we expect this ratio to increase further.   

Our mathematical formulation suggests a protraction factor for H$_2$O$_2$.
In analogy with DNA damage-repair mechanistic models, the formation of \ce{^{.}OH} and H$_2$O$_2$ are equivalent to the linear-quadratic model of the cell survival, with \ce{^{.}OH} and H$_2$O$_2$ play the role of $\alpha$ and $\beta$ indices of the radio-biological effects. Just the time scales are much shorter. Furthermore, the variable, $\Delta$, in this context, plays the role of LET as in RBE. The mathematical approach, introduced in this work, suggests bringing UHDR to the same class of LET effects, as both are correlated with the spatial compactness of ionization and excitations in the medium. Such interchangeability and equivalence of LET and UHDR effects allow both to be described by a single predictor, $\Delta$.  

We finally remark that in performing MC, incorporating inhomogeneities into the spatial and temporal track distributions must be done at the initial construction of the charged-particles phase space.
If the MC users sample a collection of the particles and generate an ensemble of the tracks from uniformly distributed selective points on a two-dimensional source, using a uniform random number generator, the generated tracks are going to be biased in favor of a uniform distribution in the medium with a maximum in the averaged inter-track separation. The outcome of such a biased MC setup does not contain inter-track correlations, simply because the MC users have started from an initial condition with no inter-particle correlations in their phase space, hence the simulations from such an MC setup do not necessarily resemble the particle distributions in the beams generated from an accelerator.

The present analysis suggests an additional index, $\Delta$ (a chemical heterogeneity index) must be added to describe FLASH-UHDR effects. $\Delta$ is the compactness of the tracks, independent of dose and dose rate. 
In other words, it is not enough to describe a beam of FLASH only by dose and dose rate. The dose rate is a single parameter and doesn't give detailed information on the spatial heterogeneity distribution and compactness of the radiation tracks and chemical reactions. The dose rate alone doesn't tell us if the chemistry is homogeneous or heterogeneous. An addition index is needed to distinguish homogeneous vs. heterogeneous chemistry. It seems $\Delta$ is an appropriate parameter to be added as a predictor for the biological effects of FLASH. Like LET, $\Delta$ is not a measurable quantity. It must be calculated. 

This highlights the importance of incorporating beam-specific parameters into the calculation of $\Delta$, as the compactness and clustering of ionization tracks are not solely a function of dose rate but are also influenced by the beam's particle type, energy, and temporal-spatial distribution. 


Additionally, this study leads us to conclude that the physics behind $\Delta$ and saturation of H$_2$O$_2$ is essentially similar to high LET saturation of RBE or scintillation quenching where a pair of electron and hole form an exciton. In our model, all of these phenomena can be described by a single parameter, the ionization compactness.
Our model bridges the gap between theory and experiment, providing a framework to understand how $\Delta$ can predict the biological effects of UHDR beams. This is particularly significant for clinical translation, as the ability to accurately describe the chemical environment generated by different radiation modalities could inform beam selection and optimization for FLASH therapy, ensuring maximum therapeutic efficacy while minimizing damage to healthy tissues. Thus, understanding the interplay between beam properties and $\Delta$ not only refines our theoretical models but also enhances their applicability to real-world clinical scenarios.

While our study attributes H$_2$O$_2$ suppression at UHDR to inter-track coupling and track clustering effects, other mechanisms may also play a role. One key alternative explanation involves oxygen effects and hypoxia-like conditions, where the FLASH effect could be linked to local oxygen levels. If oxygen depletion occurs at UHDR, it could reduce the oxidation of water, thereby lowering H$_2$O$_2$ production. Future research should focus on real-time oxygen measurements to determine whether a direct correlation exists between oxygen depletion and reduced H$_2$O$_2$ yields. Another potential factor is transient chemistry and radical reactions during the early stages of water radiolysis. Complex radical interactions, including recombination of $e^-_{\rm aq}$, \ce{^{.}OH}, and \ce{^{.}H}, may be influenced by UHDR irradiation, possibly accelerating radical recombination before H$_2$O$_2$ formation. Advanced molecular dynamics simulations could provide deeper insights into whether reaction kinetics differ significantly between UHDR and conventional dose rates. 

Our results highlight the critical role of ionization track distribution and the spatial-temporal clustering of radiation events in shaping the FLASH effect. By carefully modulating track spacing and dose delivery, FLASH therapy can be fine-tuned to maximize normal tissue protection while ensuring effective tumor control. This insight is essential for optimizing beam parameters in clinical FLASH radiotherapy. Moreover, this work bridges the gap between theory and experiment by addressing a key discrepancy: current Monte Carlo (MC)-based models often predict higher H$_2$O$_2$ production at UHDR, contradicting experimental findings. We resolve this by demonstrating that standard MC simulations assume a uniform track distribution, which fails to account for inter-track clustering in FLASH beams. This discovery validates the need for revised computational models that incorporate realistic track structures, ultimately improving dose planning accuracy in clinical FLASH trials. Moving forward, preclinical and clinical validation will be essential to confirm that lower H$_2$O$_2$ production at UHDR translates into reduced normal tissue toxicity. Additionally, a deeper understanding of track structure dynamics in FLASH therapy could pave the way for personalized treatment protocols, optimizing therapy for individual patient needs and further enhancing its therapeutic potential.

This problem continues to demand further studies for a more complete understanding.
More extensive results with different energies, particle types, their geometries, and fitting the experimental data are on the way.

\section{Appendix}
We start with the distribution function of $\theta = \overline{\nu}$ identical uncorrelated tracks leading to specific energy deposition within $z$ and $z+dz$ 
\begin{eqnarray}
F_n(z; \theta) = \sum_{\nu=0}^\infty P_\nu(\theta) f_{n,\nu}(z),
\label{App_eq1}
\end{eqnarray}
where
\begin{eqnarray}
f_{n,\nu}(z) = P_n(\mu z) \delta(z - \nu\Delta/\mu),
\label{App_eq2}
\end{eqnarray}
is the probability distribution function of exactly $n=0, 1, 2, \cdots$ ionization events generated by $\nu$ tracks. 
Note that $F_n(z; \theta)$ is a normalized distribution function, 
$1 = \sum_{n=0}^\infty\int_0^\infty dz F_n(z; \theta) = \sum_{n=0}^\infty Q_n(\theta,\Delta)$,
recalling $\sum_{\nu=0}^\infty P_\nu(\theta) = \sum_{n=0}^\infty P_n(\nu\Delta) = 1$.
The delta-function in Eq. (\ref{App_eq2}) enforces energy conservation for generation exactly-$n$ ionization events with compactness $\Delta$ from $\nu$ track. 
Using Eqs. (\ref{App_eq1}) and (\ref{App_eq2}) we find 
\begin{eqnarray}
\overline{z} &=& \sum_{n=0}^\infty\int_0^\infty dz z F_n(z; \theta)
\nonumber \\ &=&
\sum_{\nu=0}^\infty P_\nu(\theta) \sum_{n=0}^\infty \int_0^\infty dz z P_n(\mu z) \delta(z - \nu\Delta/\mu) 
\nonumber \\ &=&
\sum_{\nu=0}^\infty P_\nu(\theta)
\sum_{n=0}^\infty \left(\frac{\nu\Delta}{\mu}\right)
P_n(\nu\Delta)
\nonumber \\ &=&
\left(\frac{\Delta}{\mu}\right) \sum_{\nu=0}^\infty \nu P_\nu(\theta)
\sum_{n=0}^\infty P_n(\nu\Delta) = \left(\frac{\overline{\nu}\Delta}{\mu}\right)
\label{App_eq3_0}
\end{eqnarray}
where $\overline{\nu}=\theta$ and 
\begin{eqnarray}
\overline{z^2} &=& \sum_{n=0}^\infty\int_0^\infty dz z^2 F_n(z; \theta)
\nonumber \\ &=&
\sum_{\nu=0}^\infty P_\nu(\theta) \sum_{n=0}^\infty \int_0^\infty dz z^2 P_n(\mu z) \delta(z - \nu\Delta/\mu) 
\nonumber \\ &=&
\sum_{\nu=0}^\infty P_\nu(\theta)
\sum_{n=0}^\infty \left(\frac{\nu\Delta}{\mu}\right)^2
P_n(\nu\Delta)
\nonumber \\ &=&
\left(\frac{\Delta}{\mu}\right)^2 \sum_{\nu=0}^\infty \nu^2 P_\nu(\theta)
\sum_{n=0}^\infty P_n(\nu\Delta)
\label{App_eq3}
\end{eqnarray}
where $\sum_{n=0} P_n(\nu\Delta) = 1$, independent of $\nu$ and $\Delta$. 
As for a Poisson distribution function $\sum_{\nu=0}^\infty \nu^2 P_\nu(\theta) = \overline{\nu}(\overline{\nu} + 1)$ and $\overline{\nu} = \theta$, we obtain
\begin{eqnarray}
\overline{z^2} = \left(\frac{\Delta}{\mu}\right)^2 \overline{\nu}(\overline{\nu} + 1) = D^2 + \frac{\Delta}{\mu} D,
\label{App_eq4}
\end{eqnarray}
where $D = \theta\Delta/\mu = \overline{n}/\mu$. 
Note that as $\overline{z} = D$, Eq. (\ref{App_eq4}) can be rearranged in the following form $\Delta/\mu = \left(\overline{z^2} - \overline{z}^2\right) / \overline{z}$.
To see why $\overline{n}=\theta\Delta$, let us consider Eq. (\ref{eq1}) and set all $\gamma$'s to zero. In this limit, $\overline{n} = \mu \overline{z}$ is the trivial solution of Eq. (\ref{eq1}) after averaging over energy deposition distribution, hence $\overline{n}=\theta\Delta$ as shown in Eq. (\ref{App_eq3_0}).

\begin{figure}
\begin{center}\vspace{-0.75cm}
\includegraphics[width=1.0\linewidth]{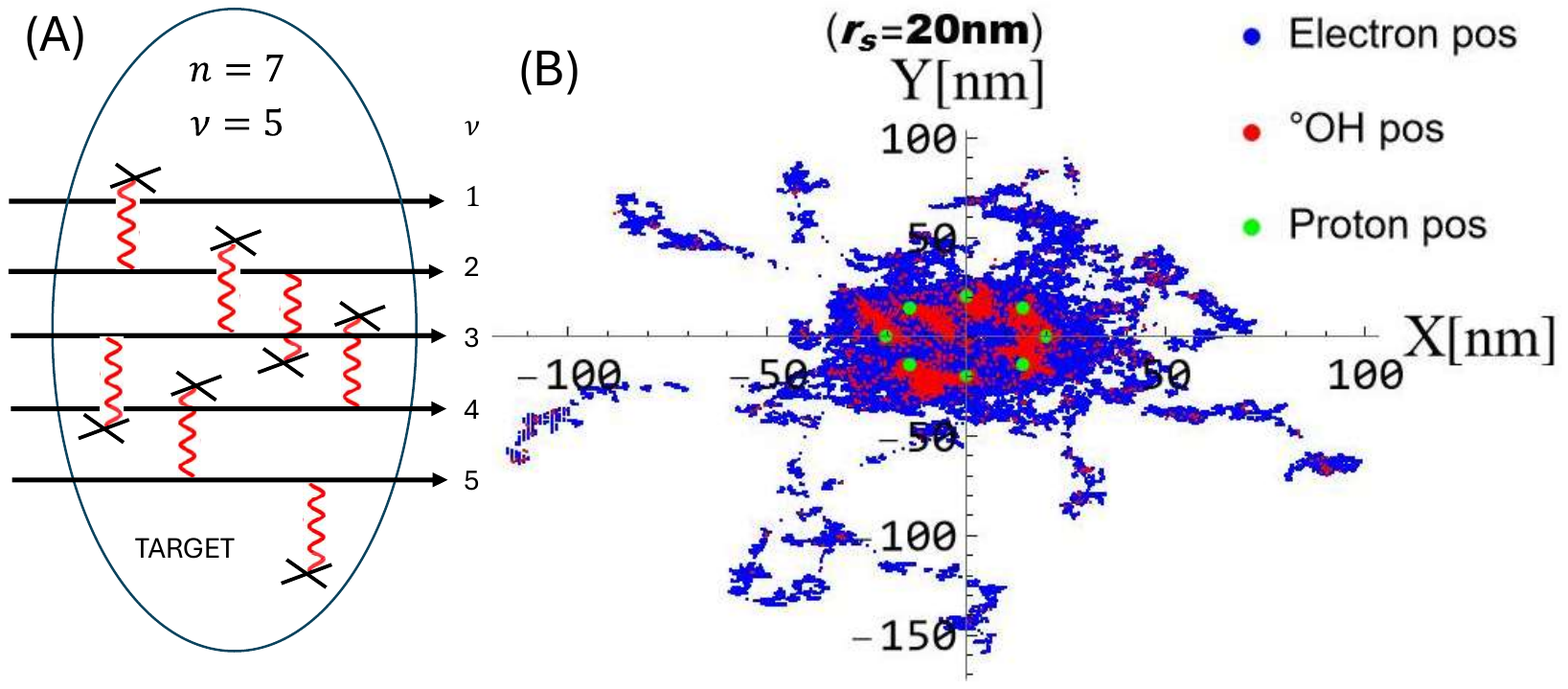} \vspace{-1.75cm} 
\noindent
\caption{
(A) Schematic view of tracks of five protons entering the medium in a bunch, resulting in seven ionization events. These five tracks, if they were well-separated, could result in $n > 7$ ionization with the vanishing chance of overlap among ion pairs within molecular spacing.
(B) Lateral distribution of ions scored by Geant4-DNA MC toolkit for eight protons with 20 nm separation, projected on XY-plane. As shown, a large number of overlaps among ion pairs may lead to double-counting of ionization events. 
}
\label{fig11}
\end{center}\vspace{-0.5cm}
\end{figure}

The present formalism allows the construction of the track-track correlations at the moment of energy deposition, as depicted in Fig. \ref{fig11}.  
To illustrate this idea, let us expand the probability distribution function (PDF) of the sum of a pair of tracks in terms of the superposition of their joint PDFs  
\begin{eqnarray}
f_{n, 1+1}(z) &=&  
\sum_{n_1=0}^\infty\int_0^\infty dz_1 \sum_{n_2=0}^\infty\int_0^\infty dz_2 f_{n_1,1; n_2,1}(z_1, z_2) \nonumber \\ &\times&
\delta_{n,n_1 + n_2} \delta(z - z_1 - z_2)
\nonumber \\ &=&
\sum_{n'=0}^\infty\int_0^\infty dz' f_{n',1; n-n',1}(z', z-z'). 
\label{Eq_App_54}
\end{eqnarray}
Here $f_{n, 1+1}(z) = f_{n, 2}(z)$ is the PDF of the sum of two tracks with $n = n_1 + n_2$ and $z = z_1 + z_2$.
When the tracks are independent, their joint PDF, $f_{n',1; n-n',1}(z', z-z')$,
factors into the convolution of the marginal PDFs: $f_{n',1; n-n',1}(z', z-z') = f_{n',1}(z') f_{n-n',1}(z-z')$ thus
\begin{eqnarray}
f_{n, 2}(z) &=& \sum_{n'=0}^\infty\int_0^\infty dz' f_{n',1}(z') f_{n-n',1}(z-z'), 
\label{Eq_App_55}
\end{eqnarray}
where $\overline{z_1 z_2} = \overline{z_1}~\overline{z_2}$, and $\Delta = (\Delta_1 + \Delta_2)/2$. 
Hence the present formulation reduces to the standard microdosimetry model developed by Rossi-Kellerer-Zaider [\onlinecite{Rossi1996:book}] and their colleagues. 

However, in a general situation when $f_{n',1; n-n',1}(z', z-z')$ is not separable, the following inequality holds, $\overline{z_1 z_2} \neq \overline{z_1}~\overline{z_2}$. Under these conditions, we can deduce $\Delta < (\Delta_1 + \Delta_2)/2$ from Eq. (\ref{Eq_App_54}), which
corresponds to the configurations of the charged particles and their tracks in the medium with a lesser number of ionizations than the algebraic superposition of each individual track due to inter-track correlations. 
As shown in Fig. \ref{fig11}, the denser the tracks, the more abundant the overlap among ion positions and the higher the overestimation of energy deposition because of double counting of ionization events.   
The details of this calculation are beyond the scope of the present study and can be postponed to our future publications, focusing only on the details of the nanoscopic energy deposition. However, below, we present a qualitative description of this inequality.   
Consider, for example, $n_1=7$ and $n_2=8$, the number of ions generated by two tracks if they were far from each other (independent tracks). In this case, $n=n_1+n_2 = 7 + 8 = 15$. If we bring these two tracks close to each other, there could be an overlap among adjacent ions that prevents the generation of some of them. In this example, the total ionization may not be essentially a sum of 7 and 8, e.g., if two ions are close to each other, only one of them may have a chance of creation, hence $n=14$. If we express $n$ as a superposition of ion counts for individual tracks, regardless of their spacing (e.g., $n=n_1+n_2$), then we find $\Delta = \overline{n}/\overline{\nu} < (\Delta_1 + \Delta_2)/2$. In our numerical example for two overlapping tracks, $\Delta = 14/2 < 15/2$.
To make this scenario more realistic, consider two tracks generated by two identical charged particles, entering the medium within nm spatial separation and ns relative time delay. 
This time interval is too short for the changes in the medium to be equilibrated and damped to the initial molecular configuration, prior to radiation, e.g.,    
the second particle enters a medium that had been already ionized by the first charged particle. 
Thus, due to the rapid changes in the nano-scale molecular environment subsequent to the generation of the first track, the second charged particle experiences a different 
electromagnetic environment [\onlinecite{Niederhaus2025:EL}]. Under such non-equilibrium conditions, the instantaneous electromagnetic couplings in the Bethe-Bloch stopping power lead to lower energy deposition due to lower absorption from the medium. 
This influences the PDF of the ionization generated by the second charged particle. 
A lower ionization density in the second track is expected in comparison with a situation if such nanoscopic changes in the electromagnetic response of the medium were absent. This results in lower net ionization from both tracks, a phenomenon known as ``ionization quenching", observed at FLASH-UHDR. 
In addition to electromagnetic changes, there are structural changes in the molecular configurations as the tracks may generate shockwaves and nano-scale bubbles with supercritical states of matter, as discussed in our recent publication [\onlinecite{Abolfath2024:EPJD}].

The simulation of such effects in MC is highly challenging. Because the adjustable parameters in the Bethe-Bloch stopping power and the electrodynamical responses of the environment must be implemented as a functional of time, including the dielectric constant, absorption coefficient, refractive index, polarization, nano-meter mass density, etc. 
Moreover, the changes to the electronic configuration of molecules, including the calculation of the electronic ionization energy of the medium, must be performed by 
time-dependent ab-initio density functional theory [\onlinecite{KC2022:SR}].
The second particle sees a new environment (e.g., in the form of a generated local plasma), and an updated Bethe-Bloch stopping power must be used. This has to be done for millions of charged particles.

\begin{figure}
\begin{center}
\includegraphics[width=1.0\linewidth]{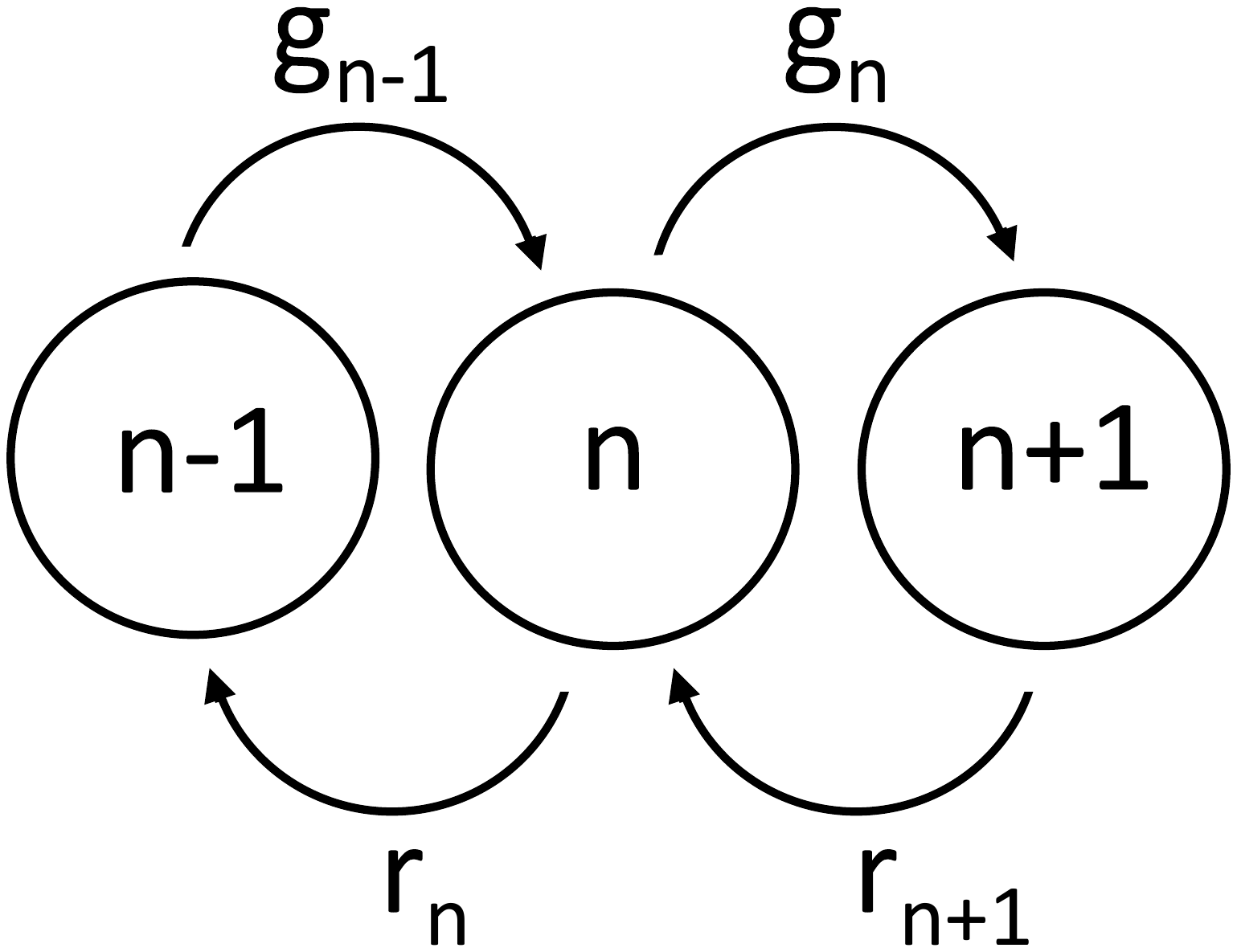} 
\noindent
\caption{
Shown schematically a typical Markov chain. 
}
\label{fig12}
\end{center}\vspace{-0.5cm}
\end{figure}

\subsection{Markov chain}
The Markov chain describes a stochastic process in time as depicted schematically in Fig. \ref{fig12}. The events at a specific time $t$ depend only on one step behind with no
long-range history from the past. As shown the transition rates $g_n$ and $r_n$ move the event forward and backward, $n \rightarrow n+1$ and
$n \rightarrow n-1$, respectively. 
The master equation associated with such sa tochastic process is given by [\onlinecite{vanKampen:Book}]
\begin{equation}
\frac{dQ_n(t)}{dt} = g_{n-1} Q_{n-1} - g_n Q_n + r_{n+1} Q_{n+1} - r_n Q_n,
\label{eq002}
\end{equation}
$Q_n$ is the normalized ($\sum_{n=0}^\infty Q_n(t)=1$) probability of occurrence of event $n$th.

It is straightforward to show the rate equation that counts for the time-evolution of $\overline{n} = \sum_{n=0}^\infty n Q_n(t)$, can be given by
\begin{eqnarray}
\frac{d\overline{n}(t)}{dt} = \sum_{n=0}^\infty (g_n - r_n) Q_n(t).
\label{eq0025A}
\end{eqnarray}
Considering 
$g_n = \mu \dot{z}$ for any $n$, and 
\begin{eqnarray}
r_{n} &=& \gamma_1 n + \gamma_2 n(n-1)/2 + \gamma_3 n(n-1)(n-2)/3! \nonumber \\ &&
+ ... + \gamma_N n(n-1)(n-2)...(n-N)/N!,
\label{eq0025Ar}
\end{eqnarray}
it is straightforward to derive Eq. (\ref{eq1}).

\begin{acknowledgements}
The authors would like to thank useful scientific communications with
Niayesh Afshordi, Martin R\"adler, Reza Taleei, Julie Lascaud, Alexander Baikalov, Stefan Bartzsch, Emil Schueler, Radhe Mohan, and Joao Seco.
\end{acknowledgements}

\noindent{\bf Authors contributions:}
RA: proposed the scientific problem, wrote the main manuscript, prepared figures, and performed mathematical derivations and computational steps.
SF: wrote the main manuscript, performed MC simulation, and prepared figures.
RA and AG: wrote the main manuscript, co-supervised this work as part of PhD thesis of SF, and provided critical feedback.
HC and MCatV: proposed the scientific problem and challenges in interpreting experimental data, performed experiments, and provided data and critical feedback.

\noindent{\bf Competing financial interest:}
The authors declare no competing financial interests.

\noindent{\bf Data Availability Statement:}
No Data is associated with the manuscript.

\noindent{\bf Corresponding Author:      $~~~~~~~~~~~~~~~~~~~~~~~~~$}\\
$^\dagger$: ramin1.abolfath@gmail.com, ramin.abolfath@howard.edu

\noindent{\bf Authors' emails:      $~~~~~~~~~~~~~~~~~~~~~~~~~$}\\
ramin.abolfath@howard.edu \\
ghasemi@guilan.ac.ir \\
$\rm{s}_{--}$fardirad@phd.guilan.ac.ir 


\end{document}